\documentclass[3p,single-column]{elsarticle}

\usepackage{lineno,hyperref}
\journal{Journal of Sound and Vibration}
\usepackage{numcompress}\biboptions{authoryear}

\usepackage{graphicx} % Required for inserting images
\usepackage[utf8]{inputenc}
\usepackage[english]{babel}
\usepackage{hyperref}
\usepackage{amsmath}
\usepackage{amsfonts}
\usepackage{comment}
\usepackage{caption}
\usepackage{subcaption}
\usepackage[normalem]{ulem} % either use this (simple) or
\usepackage{tikz}
\usepackage{pgfplots}
\usepackage{booktabs}
\usepackage{enumerate}
\usepackage{bm}
\usepackage{tabularx}  % For flexible-width columns
\usepackage{multicol}

\pgfplotsset{compat=newest}

\usepackage{amssymb}
\usepackage{tcolorbox} % For boxed layout
\usepackage{nomencl}
\usepackage{framed}
\usepackage{nomencl}
\usepackage{mdframed} % For framed environments that can span across pages
\makenomenclature

\renewcommand*\nompreamble{\begin{multicols}{2}}
\renewcommand*\nompostamble{\end{multicols}}

\usepackage{etoolbox}

\renewcommand\nomgroup[1]{%
    \vspace{5pt} % Add vertical space before each group
    }
\usepackage{amsmath}
\usepackage{amssymb}
\usepackage{tikz}
\usetikzlibrary{shapes}
\usepackage{xcolor}

\begin{document}

\begin{frontmatter}
        \title{POD-Based Sparse Stochastic Estimation of \\Wind Turbine Blade Vibrations}

        \date{January 2025}

		\author[label1,label2]{Lorenzo~Schena}
		\author[label1]{Wim~Munters}
		
		\author[label2,label3]{Jan~Helsen}

		\author[label1,label4,label5]{Miguel~ A.~ Mendez}

		\affiliation[label1]{organization={von Karman Institute for Fluid Dynamics},%Department and Organization
			city={Sint-Genesius-Rhode},
			postcode={1640}, 
			country={Belgium}}

		\affiliation[label2]{organization={Vrije Universiteit Brussel (VUB), Department of Mechanical Engineering},%Department and Organization
			city={Elsene, Brussels},
			postcode={1050}, 
			country={Belgium}}   

        \affiliation[label3]{organization={Flanders Make at VUB, BP\&M},%Department and Organization
			city={Elsene, Brussels},
			postcode={1050}, 
			country={Belgium}}  

         \affiliation[label4]{organization={Aero-Thermo-Mechanics Laboratory, Université Libre de Bruxelles},%Department and Organization
			city={Elsene, Brussels},
			postcode={1050}, 
			country={Belgium}}

        \affiliation[label5]{organization={Aerospace Engineering Research Group, Universidad Carlos III de Madrid},%Department and Organization
			city={Leganés},
			postcode={28911}, 
			country={Spain}}

\begin{abstract}
This study presents a framework for estimating the full vibrational state of wind turbine blades from sparse deflection measurements. The identification is performed in a reduced-order space obtained from a Proper Orthogonal Decomposition (POD) of high-fidelity aeroelastic simulations based on Geometrically Exact Beam Theory (GEBT). In this space, a Reduced Order Model (ROM) is constructed using a linear stochastic estimator, and further enhanced through Kalman fusion with a quasi-steady model of azimuthal dynamics driven by measured wind speed.
The performance of the proposed estimator is assessed in a synthetic environment replicating turbulent inflow and measurement noise over a wide range of operating conditions. Results demonstrate the method's ability to accurately reconstruct three-dimensional deformations and accelerations using noisy displacement and acceleration measurements at only three spatial locations. These findings highlight the potential of the proposed framework for real-time blade monitoring, optimal sensor placement, and active load control in wind turbine systems.
\end{abstract}

   
\end{frontmatter}

% Optionally, remove the automatic nomenclature title:
\renewcommand{\nomname}{}

%\begin{center} % Center the frame on the page
  %\begin{framed}
%    \textbf{List of Acronyms}\par\medskip % Manually add a title inside the frame
%    \printnomenclature
%  \end{framed}
%\end{center}

\renewcommand{\nomname}{}
\begin{center}
	\begin{framed}
		\textbf{List of Acronyms}\vspace{-0.5em}
		\printnomenclature
	\end{framed}
\end{center}

\section{Introduction}\label{s_0}   
Wind turbine blades are exposed to various sources of unsteady loads arising from rotation, turbulence, control inputs (e.g., cyclic pitch), and other environmental factors \citep{soker2013loads, schubel2012wind}. These loads, coupled with the unprecedented flexibility of modern blades, lead to large deflections and motions. Blade deformations can become large enough to change the inertial properties of the blades, altering their modal structure and resonant response depending on the operating conditions\citep{lopez-lopezDynamicInstabilityWind2020, skjoldanImplicitFloquetAnalysis2012, bottassoModelIndependentPeriodic2015, rivaPeriodicStabilityAnalysis2016}, or even azimuthal positions \citep{acar2018bend}. Moreover, large deflection can degrade aerodynamic performances \citep{larsenAeroelasticEffectsLarge}, induce harmful aeroelastic instabilities \citep{rasmussen2003present,kallesoeEffectSteadyDeflections2011}, increase aerodynamic-driven fatigue loads \citep{liu2017vibration} and intensify mode couplings induced by blade geometry (e.g., pre-bending) and the anisotropic properties of modern composite materials \citep{stablein2017modal}.

Monitoring and predicting blade behaviour is therefore critical for wind turbine operation, to guide active load mitigation strategies \citep{kragh2014sensor, cooperman2015load}, to provide advanced indicators for predictive maintenance \citep{hameed2009condition}, and to validate widespread aeroelastic tools employed by the industry and academia alike \citep{lehnhoff2020full}. Yet, high-fidelity modelling of these fluid-structure interactions requires computationally expensive numerical solvers \citep{wang2016state, li2020aerodynamic}, unsuitable for applications in which rapid feedback is of the essence.

This has driven extensive research into Low-Order Models (LOMs) and Reduced-Order Models (ROMs), which offer complementary pathways for fast predictions. On the one hand, LOMs are based on lumped descriptions that make simplifying yet reasonable assumptions \citep{pao2009tutorial, kallesoe2006low}. These models prioritize interpretability and simplicity and are generally amenable to analytical treatment, offering the simplest approach to control, stability analysis, and optimization. However, the derivation of LOMs becomes impractical as the complexity of the structural response increases\citep{volkLargeWindTurbine2020}. 
% TODO: add LOM failure reference 

In contrast, ROMs reduce the dimensionality of a high-fidelity simulation by projecting the high-dimensional state onto a reduced-dimensional space. Projection-based ROMs differ primarily in the choice of basis functions used for projection. Common choices are mode shapes obtained from linearized formulations \citep{sonderby2013low, jonkman2018full}, potentially prescribed according to the operative condition \citep{adegas2013reduced}, or general-purpose bases such as Rayleigh-Ritz \citep{branlardGeneralizedWindTurbine2024} or finite element methods (FEM) \citep{rezaeiDevelopmentReducedOrder2015}. Alternatively, hybrid linear methods could be obtained by expanding traditional bases of eigenmodes (or Linear Normal Modes, LNM), which cannot handle strong nonlinearities \citep{gozcu2020representation}, with modes such as FEM \citep{tarpoExpansionExperimentalMode2020, iliopoulos2016modal} or heuristic bases tailored by optimization \citep{gozcu2022correction}.  The nonlinear blade dynamics can also be directly addressed via the identification of the invariant manifolds of the system dynamics, described by Nonlinear Normal Modes (NNM) \citep{shaw1991non, pesheckModalReductionNonlinear2002,Touz2021, martin2023reduced}. However, inferring nonlinear manifolds from sparse measurements is more challenging, both because of the ill-posedness of the problem and the difficulty in inverting the nonlinear mapping from low to high dimension.

In monitoring applications, these reduced-order representations enable the integration of model predictions with measurements, providing computationally efficient methods for identifying and tracking a physical system state in real-time. This integration can enable (1) sparse sensing if low-dimensional measurements are used to infer the full-dimensional state; (2) virtual sensing if sparse sensing is combined with a process model that maps observables to unobserved quantities of interest (QoIs); and (3) digital twinning if the continuous stream of information is used to tailor a model to a specific system or machine.

Sparse sensing methods optimally locate sensors to retrieve the maximum amount of non-redundant information \citep{manohar2018data}, and are mostly used in the context of Operational Modal Analysis (OMA) \citep{schulze2016optimal, eichner2023optimal} and structural health monitoring \citep{ostachowicz2019optimization}. Virtual sensing applications use these measurements to infer unmeasured quantities combining measurements and a process model. Typical inference approaches are Kalman Filters (KF, \cite{welch1995introduction}) in their Augmented (AKF) formulation, to simultaneously estimate both observed (or primary) states and unmeasured (or secondary) states \cite{lourens2012augmented}. These have been used for strain estimation during so-called pull and release tests of wind turbine blades \citep{vettori2020virtual}, or the inference of tower-bottom moments \citep{branlardAugmentedKalmanFilter2020}, to give some examples. In these approaches, process models can be LOR, ROM, or purely data-driven. For example, \cite{bilbao2022virtual} used a Gaussian process and Kalman filtering to estimate tower loads,  while \citep{azzam2021development} combined neural networks with multibody models for virtual sensing of gearboxes. An example of the application of ROM as a process model for statistical inference via Kalman filtering is the work by \citep{mehrjoo2022optimal} on the use of data-driven modal analysis for optimal sensor placement to estimate the compressions of the support jacket.

The combination of a simplified model for fast prediction, enhanced or tuned by sensor data, is the essence of digital twinning \citep{wright2020tell}, which aims to derive adaptive models that can cope with the evolution of physical systems. Twinning methods vary in model complexity and tuning procedure, depending on whether LOR or ROMs are used. Examples of LOR-based twinning are provided by \cite{schena2024reinforcement} and \cite{branlard2024digital}. \cite{schena2024reinforcement} used optimal control techniques to estimate the unknown aerodynamic power curve, tracking the trajectory of the original system with a simple one-degree-of-freedom model while \cite{branlard2024digital} compared symbolic \citep{branlard2021symbolic} or linearisation routines of aeroelastic solvers \citep{jonkman2018full} to estimate tower base loads.
An example of ROM-based twinning is the work by \cite{moghadam2022online}, who combined data-driven modal analysis and inference to monitor floating drivetrains.

This work presents an approach to identify a predictive ROM of blade deformation from sparse and noisy displacement sensors. The ROM is based on a reduced set of data-driven modes identified by Proper Orthogonal Decomposition (POD, \cite{sirovich1987turbulence, sirovichOptimalLowdimensionalDynamical1990, holmes2012turbulence}) of a large data set of aeroelastic simulations in a wide range of operations. We use the extensively validated OpenFAST solver \citep{jonkman2018full} with Geometrically Exact Beam Theory (GEBT) formulation for the blade dynamics \citep{wang2017beamdyn, Reissner1973OnOL}. The POD, also known as the Karhunen-Loève decomposition \citep{Karhunen1946, loeve1977elementary}, is a standard tool in the context of dimensionality reduction in fluid dynamics \citep{berkooz1993proper,Dawson_2023, Mendez_2023}, where it is traditionally used to derive a set of optimal bases (modes) for the modal decomposition of velocity fields.
Although the POD is a linear decomposition, it neither presupposes nor necessitates that the underlying process be linear. In this regard, it is as broad in its application as a Fourier decomposition \citep{berkooz1993proper}. This flexibility makes this decomposition particularly attractive for its application to blade dynamics, in which linear and nonlinear behaviour coexist in different regions of the state space. In the context of structural mechanics, it has been shown that POD modes can coincide with LNMs if the underlying system is linear \citep{feenyInterpretingProperOrthogonal2003} and lead to an optimal linear approximation of the NNMs \citep{feenyPROPERORTHOGONALCOORDINATES2002, feenyPHYSICALINTERPRETATIONPROPER1998} if the system is nonlinear.
The proposed estimator acts in the reduced order space produced by projecting the blade displacement onto the leading POD modes and fuses, using a Kalman filter, the prediction of a quasi-steady stochastic model with real-time measurements. 

The remainder of this article is organized as follows. First, §\ref{sect:prel} introduces the terminology, notation, and reference coordinate frames used in the rest of this work, and §\ref{sect:sim_conditions} follows detailing the numerical setup. Then, §\ref{sect:srbd} presents the the mathematical formulation of the sparse reconstruction technique detailing the sensor placement and sensing processes in §\ref{sect:sparse_methods}, the formulation of the reduced-order stochastic model in §\ref{sect:qs_model} and their fusion according to the Kalman methodology. Lastly, §\ref{sect:res} overviews the results of this study, and §\ref{sect:conclusions} closes this article by summing up the main findings and overviewing new possible directions.

\section{Problem Statement}\label{sect:prel}

\begin{figure}[h!] % fix shadowed blade 
    \centering
    \includegraphics[width=.7\linewidth]{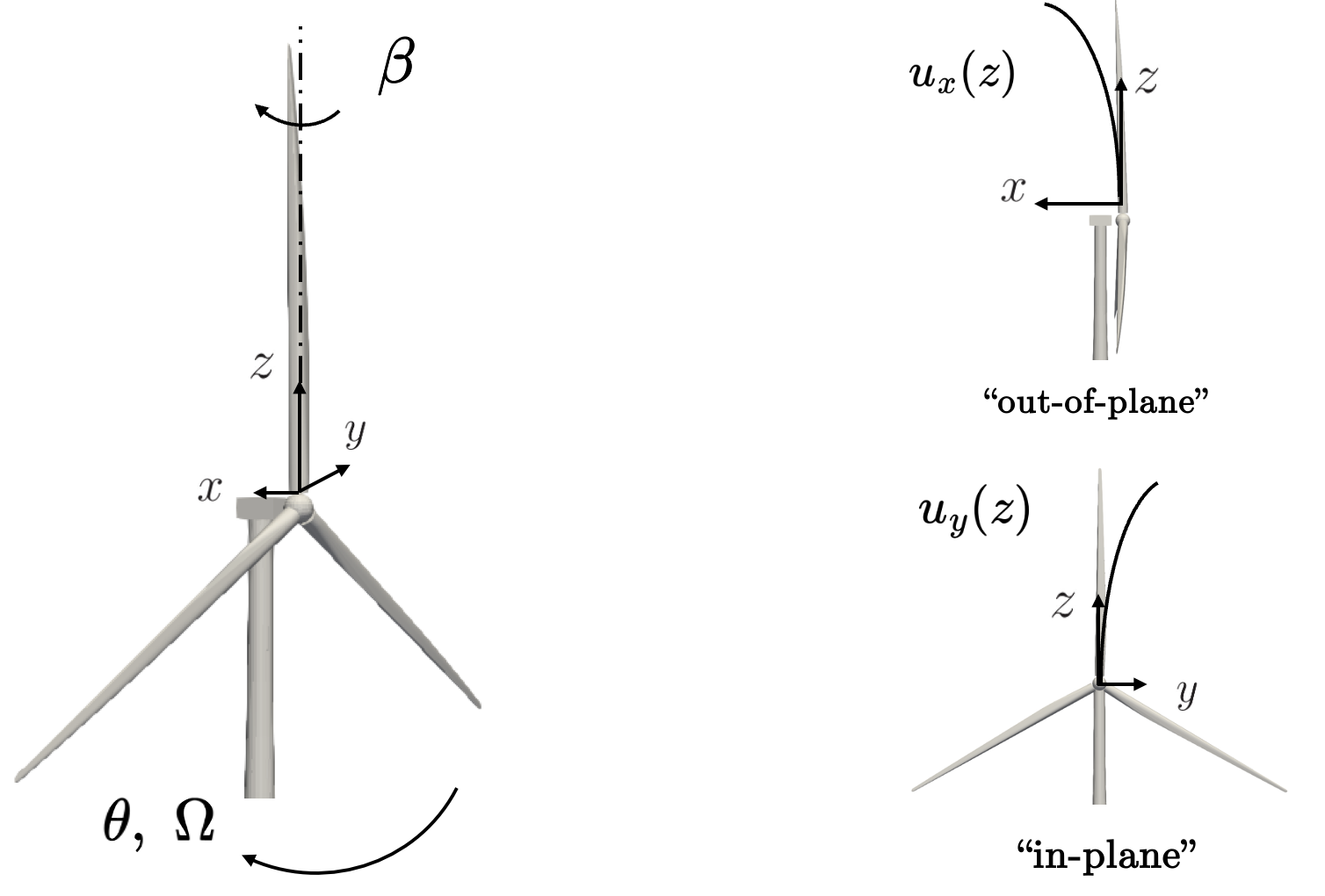}
    \caption{Blade system of reference. CAD made available by \cite{gaertner2020definition}.}
    \label{fig:reference}
\end{figure}

We denote as $\bm{u}(z, t)=(u_x(z,t),u_y(z,t),u_z(z,t))$ the three-dimensional blade deflection vector at location $z$ and time $t$, with the axis $z$ aligned with the blade axis and the entries in $\bm{u}$ corresponding to the out-of-plane deflection (or `flapwise', $u_x$), in-plane deflection (or `edgewise' $u_y$) and axial elongation ($u_z$). These quantities are illustrated in Figure 1, together with the Cartesian reference frame rotating with each blade and positioned at the intersection of the blade root and the pitch axis. Consequently, the $y$ axis is directed toward the trailing edge of the blade and is parallel to the chord in the untwisted location, while the $x$axis is orthogonal to $y$ and $z$ so that they form a right-handed coordinate system.
The blades and their coordinate system rotate with the pitch angle $\beta$ along $z$ and move in the rotor plane of an azimuthal angle $\theta$. The blade rotation is positive if anti-clockwise, as seen from upwind, leading to $\theta = 0$ for the upright position and $\theta = \pi$ for the downright position, coinciding with the tower. These conventions are illustrated in Figure \ref{fig:reference}. Note that the out-of plane deflection results in a reduction of the effective rotor disk area and thus a power loss.

For the numerical experiments in this work, measurements are simulated by sampling simulations carried out using OpenFAST \citep{openfast}, as detailed in \ref{sect:sim_conditions}. The chosen test case is the IEA-15 MW reference wind turbine (RWT) \citep{gaertner2020definition}, as a valid representative of the modern wind turbine design with flexible and slender blades. This turbine has a hub height of 150 m and $L_b$=117 m long blades. We assume that displacement measures are available with a sampling frequency of $f_s = 160$ Hz, thus on a uniform time discretization $\{\mathbf{t}_k\}^{n_t}_{k=1}=k \Delta t$ with $\Delta t=1/f_s$. This is sufficient to capture dominant blade dynamics and deflections (primarily occurring below 50 Hz for large-scale turbines) and well within the capabilities of direct or indirect displacement measurement technologies, which typically operate at significantly higher sampling rates \citep{kersemans2014ultrasonic}.% on the order of kilohertz for laser or capacitive displacement sensors and up to one order of magnitude faster for ultrasonic sensors \citet{kersemans2014ultrasonic}.

Although full-blade deflection measurements are possible using modern Bragg sensors \citep{kim2013deflection, kim2014real} or cameras \citep{lehnhoff2020full}, the common practice is to rely on less-intrusive strain sensors \citep{lee2017feasibility} to obtain deflection estimates at specific locations. In what follows, we assume that the blade under analysis is equipped with $n_P$ displacement sensors to measure all components of the displacement vector $\bm{u}$ at locations $\{\mathbf{z}_p\}_{p=1}^{n_P}$ and stored in a matrix $\bm{u}(\mathbf{z}_p, t)\in\mathbb{R}^{3\times n_P}$ at each time step. 
The measurement process is here treated as a sampling process of the continuous displacement vector polluted by a random noise:

\begin{equation}
\label{eq:sparse}
\bm{u}(\mathbf{z}_p, t) = \mathcal{H}\Big(\bm{u}(z, t) \Big) + \bm{w}(t)= \frac{1}{L_b}\int^{L_b}_0 \bm{\delta}(z - \mathbf{z}_p) \; \bm{u}(z, t)  \; \mathrm{d}z \; + \bm{W}(t)\;, \quad \text{with} \; p= 1, \dots, n_P \;,
\end{equation} where $\mathcal{H}()$ is hereinafter referred to as observation operator, $\bm{\delta} $ is the vector-valued delta function acting on each component of the displacement and 
$\bm{W}(t)\in\mathbb{R}^{n_P\times 3}$ is a vector-valued zero-mean stationary random process acting on each measurement. The random noise is assumed to be uncorrelated with the actual displacement and characterized by a set of pre-defined covariance matrices $\mathbb{E}\big(\bm{W}(t) \bm{W}(t)^\top\big) = \mathbf{\Gamma}_p\in\mathbb{R}^{3\times 3}$. We assume the random noise follows a multivariate Gaussian distribution to simplify the analysis, making model predictions Gaussian. However, as the model predictions here are limited to first- and second-order statistics, this assumption leads to no loss of generality.

\begin{comment}
\begin{equation}
    \label{eq:noise}
    \begin{cases}
        \mathbb{E}\Big(\bm{w}(t)\Big) = \mathbf{0} \;,\\
        \mathbb{E}\Big(\bm{w}(t) \bm{w}(t)^\top\Big) = \mathbf{\Gamma} \;,\\
    \end{cases}
\end{equation} 
\end{comment}

Finally, in addition to displacement measurements, it is assumed that the turbine is equipped with an anemometer to measure the wind speed $U_\infty$ at the hub height, along with sensors for azimuthal angles $\theta$ and angular velocity $\Omega$ of the rotating blades. The measured wind speed is filtered with an Exponential smoothing filter with a smoothing parameter of $\alpha=0.2$. With no loss of generality, the measurements of these quantities are assumed to be synchronously sampled with the displacement sensors. 

The scope of the sparse reconstruction is to use displacement samples at $n_p$ locations to reconstruct the ``full state'' of the blade displacement. By ``full state'', here we mean a displacement measurement in a number of locations $n_s\gg n_p$. Regression or interpolation methods could be employed to build a continuous representation of the displacement field, but such extensions are left to future work. To facilitate the assessment of the reconstruction accuracy, the full state estimation is given at the same points at which the data from numerical simulations are available. 

\section{Wind Turbine aeroelastic simulations with OpenFAST}\label{sect:sim_conditions} 

\begin{figure}
    \centering
    \includegraphics[width=1.05\linewidth]{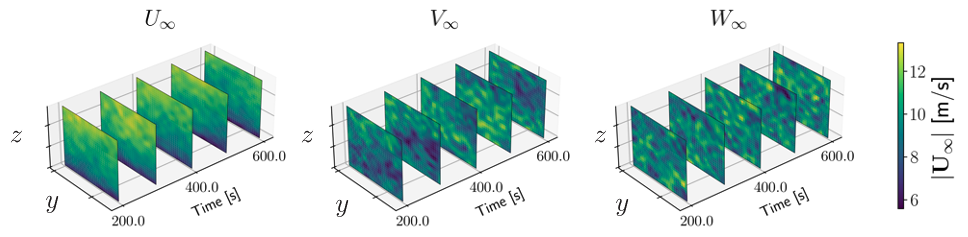}
    \caption{Example of free stream velocity $\mathbf{U}_\infty = (U_\infty, V_\infty, W_\infty)$ generated with Turbsim, for a mean horizontal wind speed of $\bar{U}_\infty = 10.6$ m/s and TI=5\%.}
    \label{fig:wind_speeds}
\end{figure}

We employ OpenFAST 3.5.3 with BeamDyn \citep{wang2017beamdyn} to describe the blade's three-dimensional non-linear elastic deformation via the geometrically exact beam theory (GEBT) formulation,and we negelect tower motions, e.g. the tower is assumed to be infinitely rigid. The blade is spatially discretized in Legendre spectral finite elements (LSFE) with quadratic convergence properties. This solver can capture coupled non-linear motions with arbitrary levels of deflections and rotations, thus relaxing the usual assumption of small deflections in the Euler-Bernoulli theory and enabling the study of torsional motions. We consider polynomials of the 14th order. Aerodynamic loads are calculated with the AeroDyn v module AeroDyn v. 15 \citep{moriarty2005aerodyn, jonkman2015aerodyn} that in our setup features a Dynamic BEM Theory (DBEMT) formulation. This model provides time-dependent corrections directly within the momentum equations, addressing the lagging effects in the unsteady aerodynamic response. In addition, we account for the dynamic stall \citep{beddoes1987near} using a Beddoes-Leishman model, as it has been shown to directly affect the damping of the flap-wise deflection modes and, ultimately, the stability of the rotating blade \citep{branlard2022dynamic}. The aerodynamic forcing is then completed by the inclusion of the tower shadowing, i.e., a localized velocity deficit in the proximity of the tower, that is modeled as the potential flow effect. 

The turbine is controlled using the Reference Open Source Controller (ROSCO) \citep{abbas2022reference}. In below-rated conditions (i.e., $\bar{U}_\infty \le 10.6$ m/s), the pitch angle is fixed at $\beta=0^\circ$ and the resistive generator torque is regulated to track the optimal tip-speed ratio, $TSR = \Omega R/U_\infty = TSR^\ast$, that results in maximum power harvesting. In above-rated conditions, the generator torque is held constant and the pitch control loop adjusts $\beta$ to reduce structural loads while keeping power output constant. In this study, we employ collective pitch control, applying the same angle to all blades.

The operative conditions studied span the entire operative range $U_\infty \in [4, 17.2]$ m/s with steps of $\Delta \bar{U}_\infty = 2.2 $ m/s, for a total of 7 operative points. The simulation time is $T = 500$ s, of which the first 100 s are discarded to neglect the initial transient time. For each operative point, we consider three different levels of turbulence intensities, specifically  TI = $[5, 10, 15] \%$, to provide a more diverse set of excitations. The Kaimal isotropic spectrum is used to generate synthetic turbulence to consider larger cyclic loading \citep{nybo2022sensitivity}, and we generate three fields per operative point using different initial random seeds. Turbulent wind fields are generated with TurbSim software \citep{jonkman2009turbsim}, and Figure \ref{fig:wind_speeds} shows an example wind field for a mean horizontal wind speed $\bar{U}_\infty=10.6$ m/s at the hub height, with turbulence intensity TI =5\%.

\section{Kalman-based Fusion of Sparse Sensor Data and Fourier Models in Reduced POD Space}\label{sect:srbd}

The proposed estimator is rooted in the modal decomposition of the blade displacement field. Denoting as $\Phi(z) = [\bm{\phi}_1(z), \bm{\phi}_2(z), \dots, \bm{\phi}_N(z)]$ the set of $N$ vector-valued continuous basis functions in $z\in[0, L_b]$, with $\bm{\phi}_n(z)\in\mathbb{R}^{3}$, 
and as $\bm{a}(t) = [a_1(t), a_2(t), \dots, a_N(t)]^\top\in\mathbb{R}^{N\times 1}$ the set of associated temporal evolutions, with $a_n(t)\in\mathbb{R}$, the generic modal decomposition reads   

\begin{equation}
    \label{eq:u_disp_modal_generic}
    \bm{u}(z, t) \approx \sum_{n=1}^N \bm{\phi}_n (z) \; a_n(t)= \Phi(z) \bm{a}(t)^\top \;.
\end{equation} 

For a complete set of basis functions $\Phi(z)$ that span the function space of interest and satisfy the boundary conditions, one expects an exact representation of the displacement field (and hence the approximation in \eqref{eq:u_disp_modal_generic} becomes equality) as $N\rightarrow \infty$. Assume that the spatial basis is orthogonal with respect to the continuous inner product $\langle\cdot,\cdot\rangle$

\begin{equation}
    \label{eq:inner}
    \langle \bm{v}_1(x), \bm{v}_2(x) \rangle = \frac{1}{L_b} \int^{{L_b}}_0 \bm{v}^\dagger_2(x) \bm{v}_1(x) \; \mathrm{d}z \;, 
\end{equation} with $\bm{v}_1(x), \bm{v}_2(x)$ two generic vector-valued functions and  $(\cdot)^\dagger$ denoting conjugate transposition, the time evolution of the vector $ \bm{u}(z, t)$ along the $n$-th element of the spatial basis can be written as 

\begin{equation}
    \label{eq:a_t_modal_proj}
    a_n(t) = \langle \bm{u}(x, t) \;, \bm{\phi}_n(z)\rangle \;.
\end{equation} 

In the following, we denote by $\bm{a}(t)=[a_1(t),a_2(t),\dots,a_N(t)]=\langle \bm{u}(z, t), \Phi(z) \rangle\in \mathbb{R}^{N}$ the projection over the set of basis elements, thus providing a $N$-dimensional representation of the displacement evolution. Although the proposed approach could be generalized to any set of basis orthogonal according to \eqref{eq:inner}, this work focuses on Proper Orthogonal Decomposition (POD), built from a large numerical database of blade deformation dynamics. This data-driven decomposition is optimal \citep{berkooz1993proper,Dawson_2023} with respect to the inner product \eqref{eq:inner} in the sense that the error of any approximation based on $N$ modes 

\begin{equation}
\label{eq:error_POD}
\bm{e}_N(z,t)=\bm{u}(z, t)-\sum_{n=1}^N \bm{\phi}_n (z) \; a_n(t)\;.
\end{equation} has the least $L_2$ error $\langle \bm{e}_N(z,t),\bm{e}_N(z,t)\rangle$ in space and within an observation time $T_o$. The POD modes are defined as the solution of the continuous eigenvalue problem 

\begin{equation}
    \label{eq:c_eig_prob}
    \int^{L_b}_0 \bm{C}(z, z') \; \bm{\phi}_n (z') \; \mathrm{d}z = \lambda_n \; \bm{\phi}_n(z) \;,
\end{equation} where $\bm{C}(z, z')\in\mathbb{R}^{3\times 3}$ is the autocorrelation tensor of the field, defined as
\begin{equation}
    \label{eq:C_tensor_def}
    \bm{C}(z, z') = \frac{1}{{T_o}}\int^{T_o}_0 \bm{u}(z, t) \; \otimes \; \bm{u}(z',t) \; \mathrm{d}t \;,
\end{equation} with $\lambda_n>0$ the eigenvalue associated to each mode and $\otimes$ the outer product between vectors. The positive definiteness of the operator $\bm{C}(z,z')$ ensures that all eigenvalues are positive and can be used to rank the associated modes by the level of importance, i.e. by their contribution to the approximation \eqref{eq:u_disp_modal_generic}. When the data is available on a uniform grid of points $\mathbf{z}_i=i\Delta z$ and uniform sampling interval $\mathbf{t}_k=k\Delta t$, the continuous eigenvalue problem can be naturally converted into a matrix eigenvalue problem. In this work, the POD was computed using the open-source Python package MODULO by \cite{poletti2024modulo}. 

This work proposes a reduced-order model for dynamics in reduced space $\bm{a}(t)$, constructed by combining two sources of information. These include the dynamics inferred from sparse sensor measurements, denoted as $\bm{a}_\bullet(t)$, and the prediction based on a quasi-steady azimuthal deflections model, denoted as $\bar{\bm{a}}_\circ(t)$. Denoting as $\mathbf{\Sigma}_\bullet(t)$ and $\mathbf{\Sigma}_\circ(t)$ the covariance matrices of these two sources, the mean prediction and the associated covariance function of the optimal fusion are given by

\begin{align}
    \label{eq:kalman_as_1}
    \hat{\bm{a}}(t) &= \bar{\bm{a}}_\circ(t)  + \mathbf{K}\big(\bar{\bm{a}}_\bullet(t) - \bar{\bm{a}}_\circ(t) \big)\;, \\
    \label{eq:kalman_variance}
    \hat{\mathbf{\Sigma}}(t) &= \big(\mathbf{I} - \mathbf{K}\big)\mathbf{\Sigma}_\circ(t)\;,
\end{align} where 

\begin{equation}
    \label{eq:kalman_gain}
    \mathbf{K} = \mathbf{\Sigma}_\circ\big(\mathbf{\Sigma}_\circ + \mathbf{\Sigma}_\bullet\big)^{-1} \;,
\end{equation} is the optimal Kalman gain. This Kalman fusion is optimal in the sense that the covariance $\hat{\mathbf{\Sigma}}(t)$ obtained by the combination has the smallest possible trace. As a result, the combination weights the contribution of the fused predictions according to their uncertainties \citep{stengel1994optimal, pei2019elementary}. The remainder of this section details the computation of the reduced-order estimate from sparse sensors ($\bm{a}_\bullet(t),\mathbf{\Sigma}_\bullet(t)$) in §\ref{sect:sparse_methods} and the quasi-steady stochastic deflections model ($\bm{a}_\circ(t),\mathbf{\Sigma}_\circ(t)$) in §\ref{sect:qs_model}. 

\subsection{Stochastic Estimation of Reduced-Order Dynamics from Sparse Sensor Data}\label{sect:sparse_methods}

A Linear Stochastic Estimator (LSE) was used to infer the conditional probability distribution of the $N$ modal amplitudes from a set of $n_P$ measurements. In our approach, we use $n_P = N$. Denoting as $\bm{\Phi}(\mathbf{z}_i) = \big[\bm{\phi}_1(\mathbf{z}_i), \bm{\phi}_2(\mathbf{z}_i), \dots, \bm{\phi}_{n_P}(\mathbf{z}_i) \big] \in \mathbb{R}^{n_S \times n_P}$ the discrete set of $n_P$ bases obtained by sampling $\bm{\Phi}(z)$ at points $\{\mathbf{z}_i\}^{n_z}_{i=1}$ and $n_S = 3n_z$ the number of total sampled scalar values from a three-dimensional field, the $n_P$ sensor locations were determined with a `greedy' strategy, i.e. performing the rank revealing QR factorization \citep{businger1965linear, strang2000linear} of the transposed basis $\bm{\Phi}(\mathbf{z}_i)^\top$.

This is a well-established approach in operational modal analysis \citep{schedlinski1996approach, schulze2016optimal} and sparse reconstruction \citep{manohar2018data}, defining optimal sensor locations as those that minimize the degree of correlation between the signals available at those locations. Writing the rank-revealing QR factorization of $\bm{\Phi}(\mathbf{z})^\top$ as 

\begin{equation}
    \label{eq:qr_fact_1}
    \bm{\Phi}{(\mathbf{z})}^\top \; \mathbf{P} = \mathbf{Q} \; \mathbf{R}\,,
\end{equation} where $\mathbf{P} \in \mathbb{R}^{n_S \times n_S}$ is the permutation matrix, $\mathbf{Q} \in \mathbb{R}^{n_P \times n_P}$ an orthogonal basis and $\mathbf{R} \in \mathbb{R}^{n_P \times n_S}$ is an upper trapezoidal matrix, the diagonal entries $(\mathbf{R}_{i+1, i+1})^{n_S}_{i = 1}$ give a metric of linear dependency of the columns of the transposed modal matrix, i.e., the linear dependency of modal information for a given location. The permutation matrix $\mathbf{P}$ sorts the diagonal elements of $\mathbf{R}$ in descending order so that the first column $n_P$ entries are the most linearly independent, in the sense that these span the largest volume in $\mathbb{R}^{n_P}$. Accordingly, the optimal sensor locations are chosen from the first $n_P$ entries of $\mathbf{P}$.

Using the optimal locations $\{\mathbf{z}_p\}^{n_P}_{p=1}$ in the observation operator \eqref{eq:sparse}, it is possible to introduce the projection of the observed deformations $\mathcal{H}(\bm{u}(z, t))$ on the observed bases $\mathcal{H}(\mathbf{\Phi}(z))$ as 

\begin{equation}
\label{eq:sparse_inner}
    {\bm{a}}_\bullet(t) = \big\langle \mathcal{H} \big(\bm{u}(z, t)\big), \mathcal{H}\big(\Phi(z)\big)\big\rangle \;.
\end{equation} At the limit $\mathbf{z}_p\rightarrow z$, the projection of observed quantities tends to the continuous projection in \eqref{eq:a_t_modal_proj}. 
% introduce gamma and sima from covariance 

Since the observation operator is linearly affine with respect to the modal amplitudes, it is possible to relate the covariance matrices in the measurement noise to the covariance matrix of the modal amplitudes. The modal amplitudes inferred from the sensor measurements becomes

\begin{equation}
    \label{eq:multivariate_meas}
    \bm{a}_\bullet(t) \sim \mathcal{N}\bigl( \bar{\bm{a}}_\bullet(t), \mathbf{\Sigma}_\bullet\bigr)\,,
\end{equation} where $\bar{\bm{a}}_\bullet(t)=\big\langle \mathcal{H} \big(\bar{\bm{u}}(z, t)\big), \mathcal{H}\big(\Phi(z)\big)\big\rangle $ is the projection of the average displacement signal, with $\overline{\bm{x}}=\mathbb{E}(\bm{x}(t))$ the ensemble average operator and $\mathbf{\Sigma}_\bullet = \mathbf{\Phi}(\mathbf{z}_i)^\top \; \mathbf{\Gamma} \; \mathbf{\Phi}(\mathbf{z}_i)\in\mathbb{R}^{n_P\times n_P}$, with $ \mathbf{\Gamma}\in\mathbb{R}^{n_S\times n_S}$ the covariance matrix assembled from the covariance matrices $\mathbf{\Gamma}_p\in\mathbb{R}^{3\times 3}$ at each of the measurement location.

% \subsection{Quasi-Steady Stochastic Fourier Models of Reduced-Order Azimuthal Blade Motions}\label{sect:qs_model}
\subsection{Azimuthally-Periodic Stochastic ROM of Blade Motions}\label{sect:qs_model}

The quasi-steady stochastic model of the blade deflection takes advantage of the azimuthal symmetry and the loading periodicity to relate the time coordinate to the angular position, hence transforming the modal time-evolution $\bm{a}(t)$ to an azimuth-based representation $\bm{a}(\theta(t))$. This model acts as a regularization for the reduced order model. The rotor plane is partitioned into $n_\theta=72$ uniform sectors to provide a 5 deg resolution. For each of the three investigated values of turbulence intensity TI$_l$, with $l\in [1,2,3]$, each bin of the filtered hub-height wind speed ($\bar{U}_{\infty,m}$), with $m\in [1,\dots 7]$, and each angular bin $\delta \theta_i$ with $i\in[0,\dots n_\theta]$, the $N$ binned modal coefficients are stored into a large snapshot matrix $\mathbf{A}_{\theta}(\delta \theta_{ i},U_{\infty,m},\mbox{TI}_l)\in\mathbb{R}^{N\times n_{a,\theta}}$ where $n_{a,\theta}$ is the number of deformation profiles available within a given bin triplet $n,m,l$. 

The mean modal amplitude and covariance matrix for each triplet is computed as 

\begin{equation}
\label{azimutal}
\overline{\bm{a}}_{\theta}(\delta \theta_n,U_{\infty,m},\mbox{TI}_l)=\frac{1}{n_{a,\theta}} \sum^{n_{a,\theta}}_{j=1}\mathbf{A}_{\theta}[:,j] \quad \mbox{and}\quad \bm{\Sigma}_\theta(\Delta \theta_n,U_{\infty,m},\mbox{TI}_l)=\frac{1}{n_{a,\theta}}\mathbf{A}_{\theta}\,\mathbf{A}_{\theta}^\top\,.
\end{equation}

As a way to prescribe a continuous functional dependency of the model from the azimuthal position and, at the same time, provide a smooth and continuous model in $\theta$, all entries of the mean coefficient vector and all entries in the covariance matrix are represented using a truncated Fourier series. Thus, for a given azimuthal location $\theta$ and environmental conditions $\Delta_{m,l} = (\bar{U}_{\infty, m}, TI_l)$, the series corresponding to the generic entry $f(\theta;\Delta_{l,m})$ of the mean and covariance models are written as

\begin{equation}
    \label{eq:fourier_series}
    f(\theta, \Delta_{l,m}) = c_{0, \Delta}(\theta) + \sum_{k=1}^{n_F} \bigg(c_{k, n}(\Delta_{l,m}) \cos(k\theta) + s_{k, n}(\Delta_{l,m}) \sin(k\theta) \bigg)\;,
\end{equation} in which $n_F=6$ and all coefficients of the expansions are obtained via linear regression from the data available at the bins $\delta \theta_i$.
The resulting stochastic azimuthal model completely defines the modal coefficients as a Gaussian process of the blade position $\theta$ and conditioned on the environmental parameters $\Delta_{l,m}$:

\begin{equation}
    \label{eq:qs_model_eq_def}
    \bm{a}_\circ (\theta ; \Delta_{l,m}) \sim \mathcal{N}\Big(\bar{\bm{a}}_\circ(\theta; \Delta_{l,m}),\; \mathbf{\Sigma}_\circ(\theta; \Delta_{l,m})\Big) \;.
\end{equation} 

This model is left discontinuous in the space of operating parameters, using a simple interpolation between these values not available in the training set.

\begin{figure}[t!]
    % First Row (Side-by-Side)
    \begin{subfigure}[b]{.53\linewidth}
         \includegraphics[width=\linewidth]{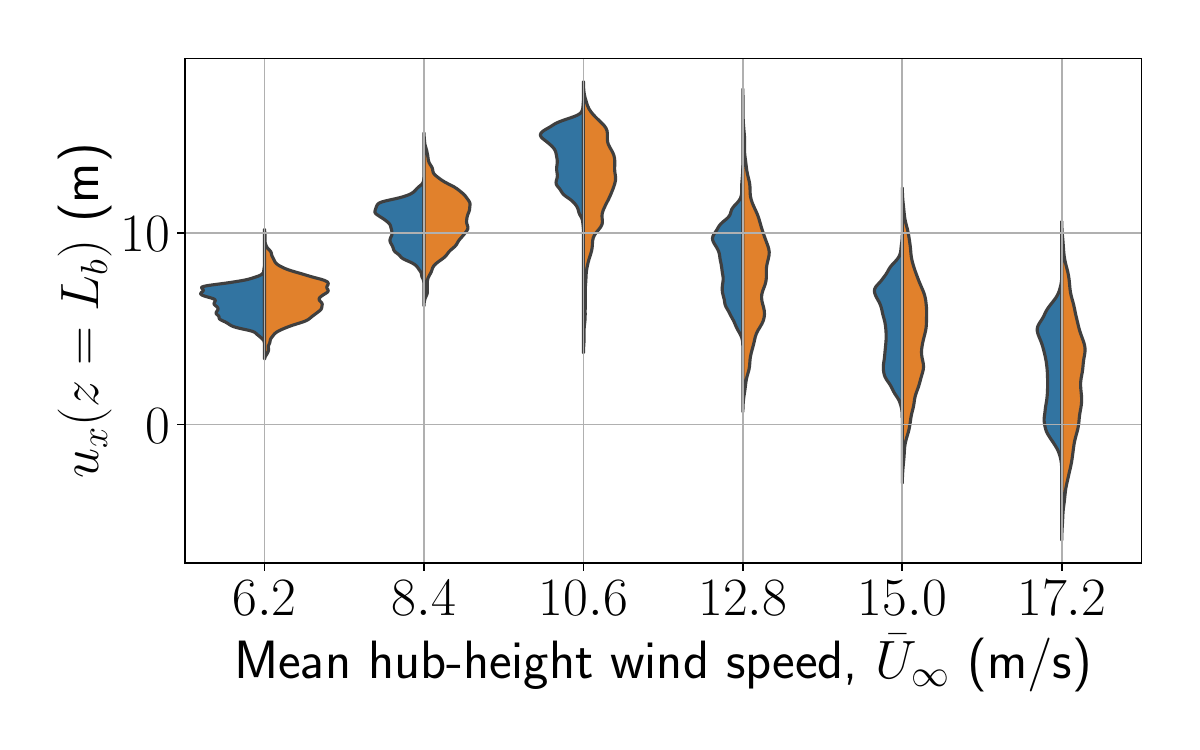}
         \caption{}
         \label{fig:violin_deflections_flapwise}
    \end{subfigure}
    \begin{subfigure}[b]{0.49\linewidth}
         % \centering
         \includegraphics[width=\linewidth]{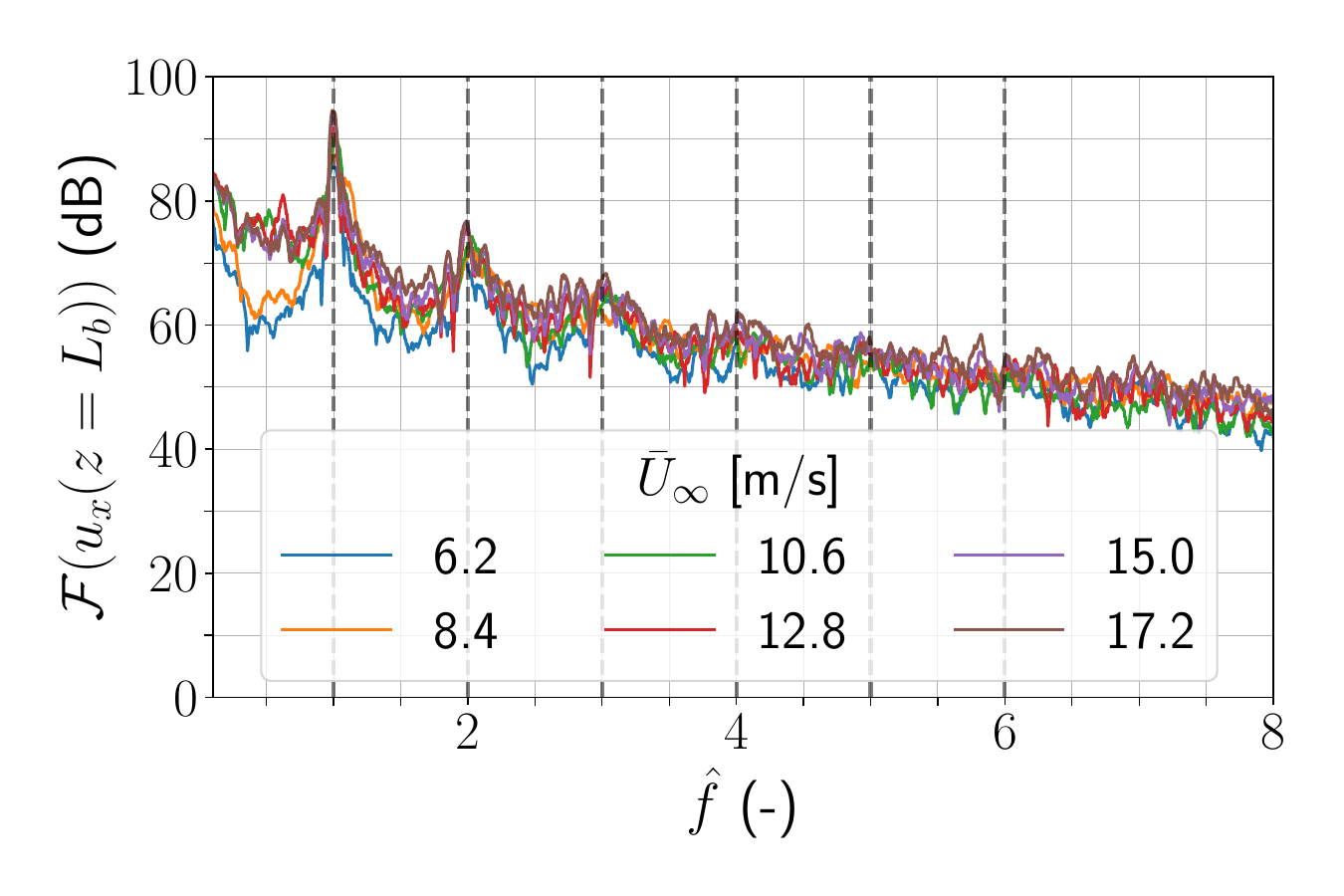}
         \caption{}
         \label{fig:psd_tip_x_defl}
     \end{subfigure}
    \begin{subfigure}[b]{.53\linewidth}
         \includegraphics[width=\linewidth]{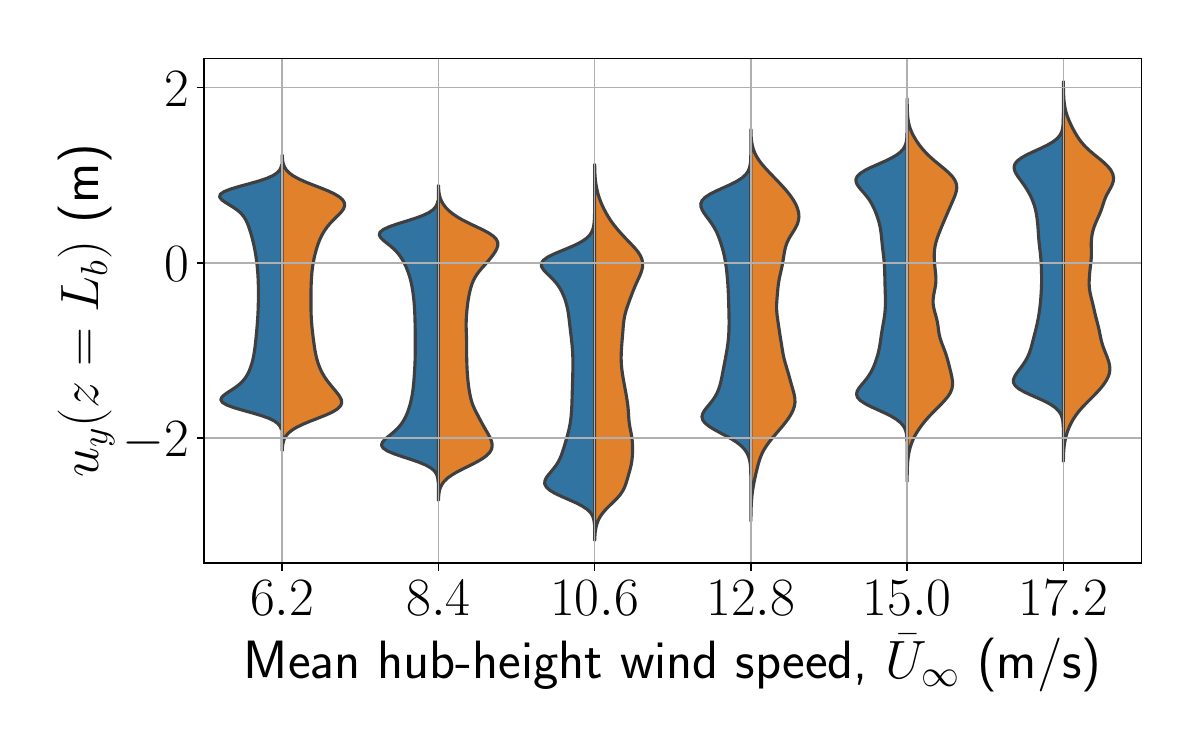}
         \caption{}
         \label{fig:violin_deflections_edgewise}
    \end{subfigure}
    \begin{subfigure}[b]{0.49\linewidth}
         % \centering
         \includegraphics[width=\linewidth]{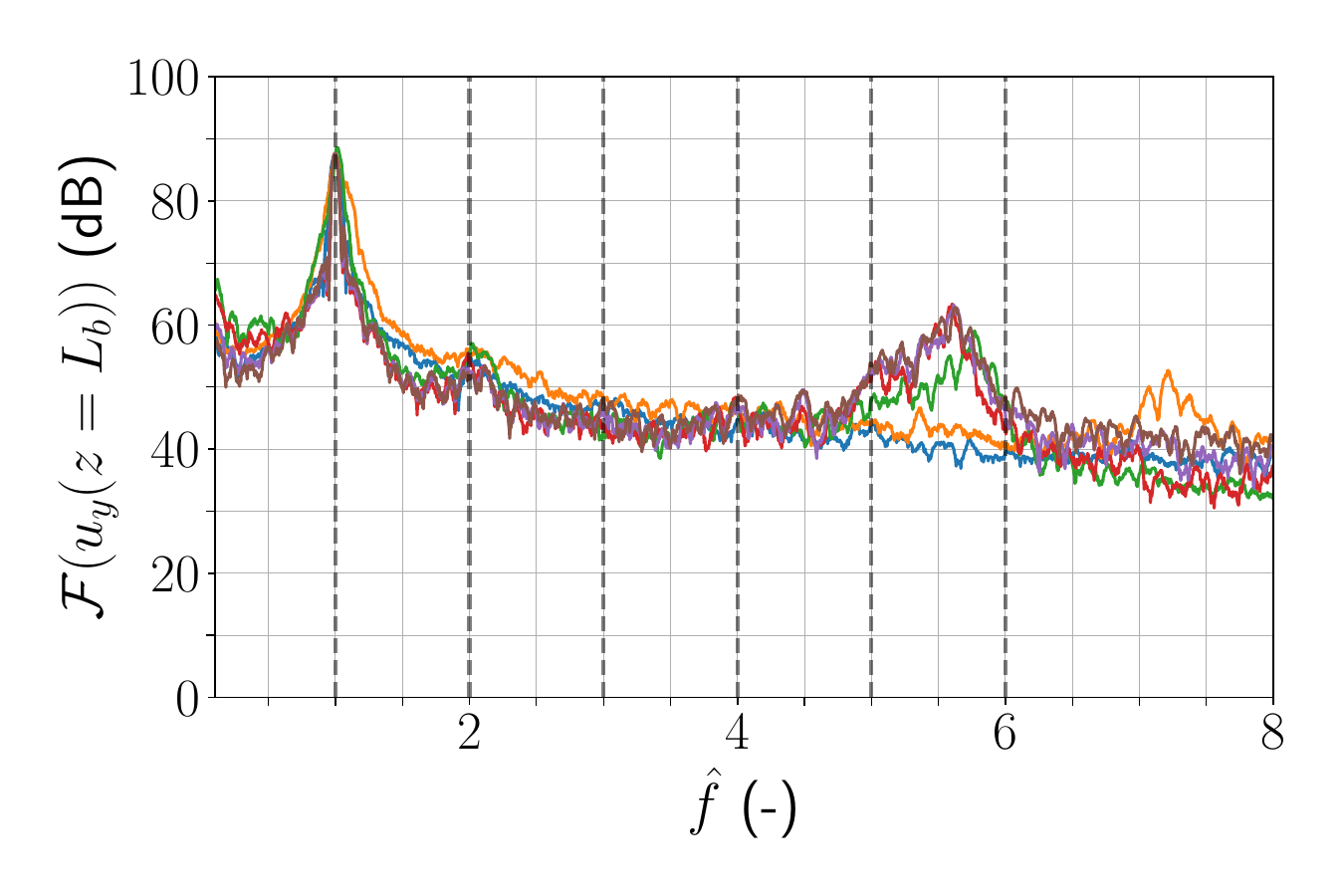}
         \caption{}
         \label{fig:psd_tip_y_defl}
     \end{subfigure}
    % Second Row (Single Centered Plot)
    \begin{subfigure}[b]{.53\linewidth}
         \centering
         \includegraphics[width=\linewidth]{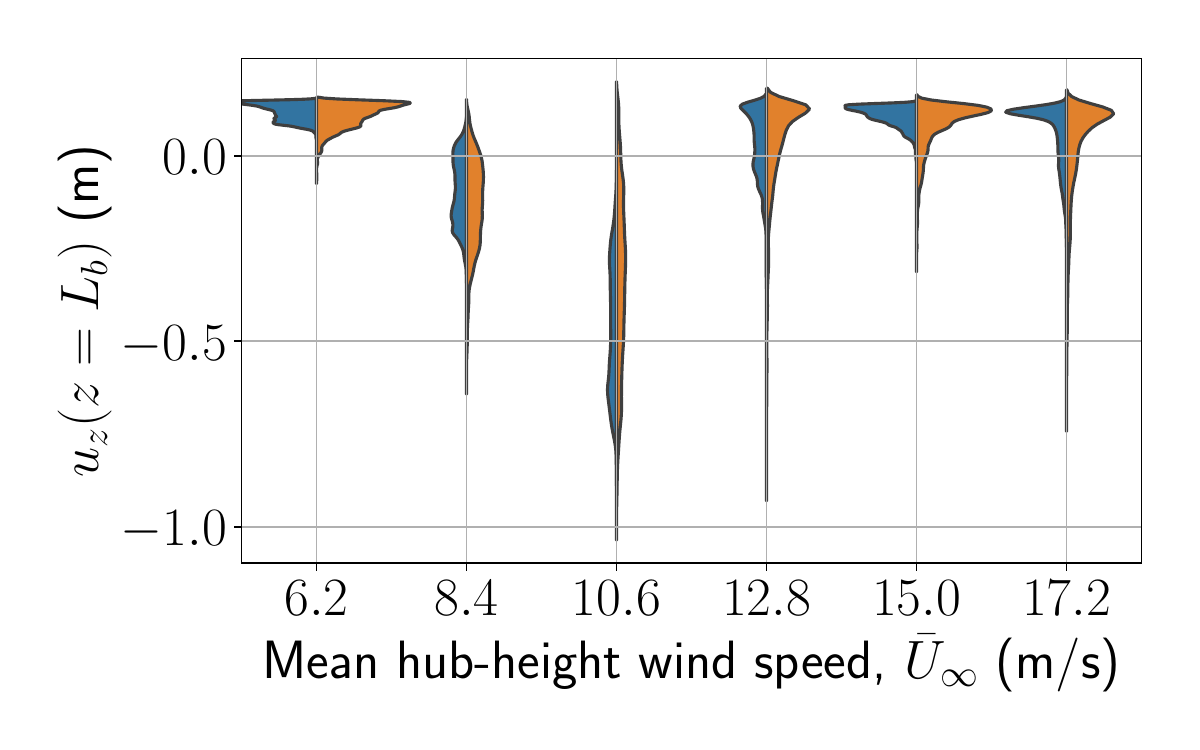}
         \caption{}
         \label{fig:violin_deflections_axial}
    \end{subfigure}
    \begin{subfigure}[b]{0.49\linewidth}
         % \centering
         \includegraphics[width=\linewidth]{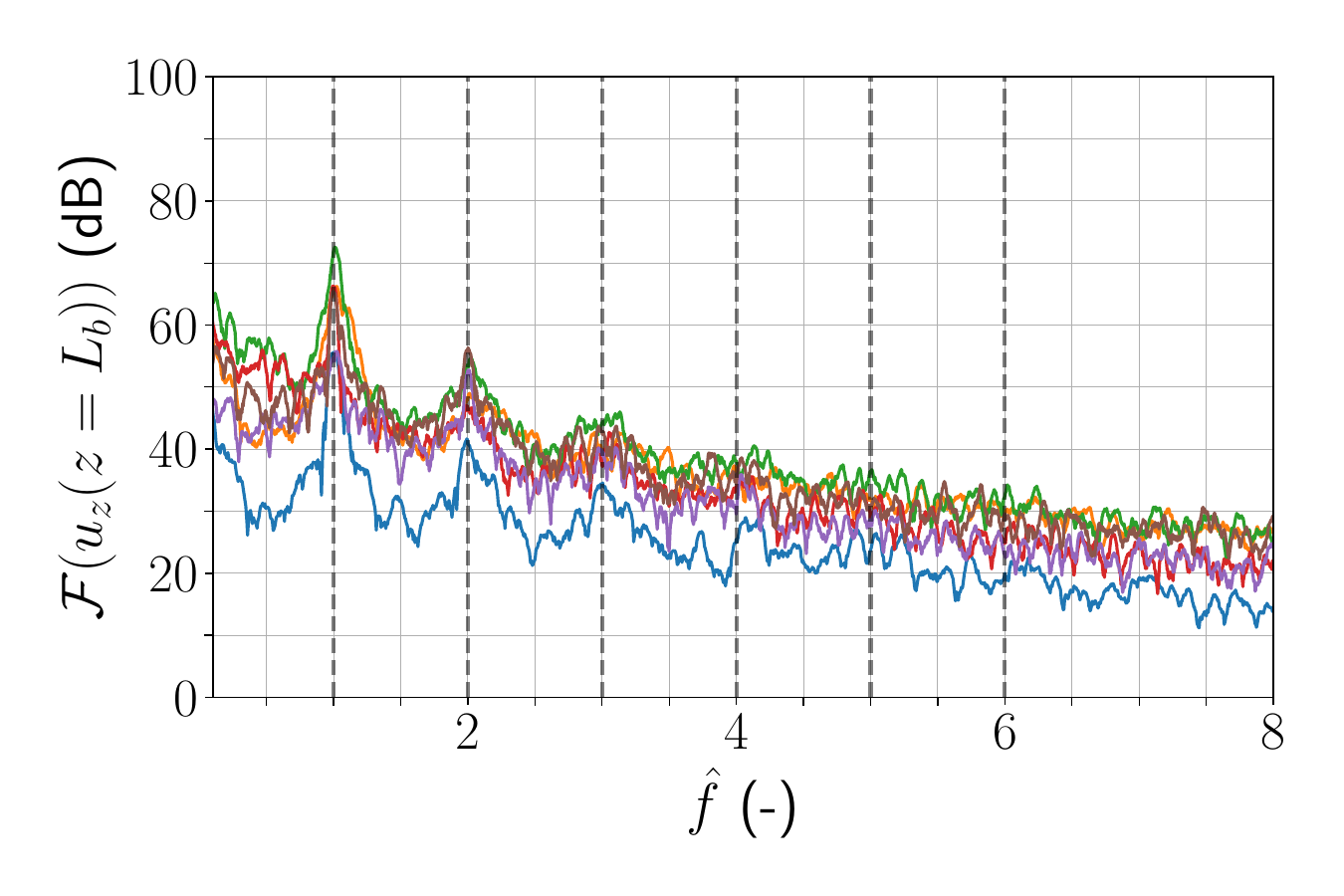}
         \caption{}
         \label{fig:psd_tip_z_defl}
     \end{subfigure}
    \caption{Analysis of the influence of the operative condition (wind speed, control) over the amplitude and frequency characteristics of $\bm{u}(z=L_b)$. The left column shows violin plots of blade tip displacement as a function of mean hub-height wind speed ($\bar{U}_\infty$), showed for TI=5\% (blue distributions) and TI=15\% (orange distributions). The right panels show their respective PSD grouped for all turbulence intensities. The raw spectra are smoothed with a Savgol filter using a window size of 33 and a third-order polynomial. The frequency axis is normalized with the 1P rotor frequency, i.e. $\hat{f}=f/f_{1P}$, and dashed vertical lines highlight the first $nP$ harmonics with $n=6$. Each row corresponds to a direction of displacement: the first row analyses the flapwise ($u_x$), the second row addresses the edgewise ($u_y$) and the last illustrates the axial response ($u_z$).}
    
    % flapwise response $u_x$ is analysed in Figures \ref{fig:violin_deflections_flapwise} - \ref{fig:psd_tip_x_defl}, the edgewise dynamics $u_y$ in Figure \ref{fig:violin_deflections_edgewise} - \ref{fig:psd_tip_y_defl}, while the axial behaviour $u_z$ is presented in Figures \ref{fig:violin_deflections_axial} - \ref{fig:psd_tip_z_defl}.
    % }
    \label{fig:tip_defl_u_ti}
\end{figure}

\section{Results}\label{sect:res}
This section is divided into four subsections. First, §\ref{res_sec_1} overviews the main characteristics of blade dynamics in the operative conditions considered, in the context of this work. §\ref{res_sec_2} presents the result of the POD on the vibration data of the blade, analysing the shape and dynamics of the identified modes and comparing them against the fundamental modes of vibration of the cantilevered beam. In §\ref{res_sec_3} we analyse the reduced-order blade dynamics in the azimuthal plane, and we show the stochastic Fourier ROM. Finally, §\ref{res_sec_4} describes the optimal sensor placement and the POD-based full-field deflection estimation.

\subsection{Blade Deformation Dynamics}\label{res_sec_1}
The salient dynamic characteristics of the blade tip dynamics $\bm{u}(z=L_b)$ are illustrated in Figure \ref{fig:tip_defl_u_ti}. The left-hand side of this figure shows the distributions of blade tip displacements relative to unloaded conditions as a function of the mean hub-height wind speed and turbulence intensity. The right-hand side complements these distributions illustrating the frequency content of the displacements, normalised against to the 1P rotor frequency. The maximum loading is met at rated conditions $\bar{U}_\infty \le 10.6 $ m/s, where the turbine blades operate at the maximum angle of attack. In this condition, the relative tip displacement reaches $u_x=15$ m, $u_y=-2.2$ m and $u_y=-0.7$ m.  The flapwise deformation, shown in Figure \ref{fig:violin_deflections_flapwise}, presents a bi-modal distribution due to the combination of wind shear and tower shadow effects. This cyclic effect induces larger deflections when the blade is upright and lower loads near the tower, and it exhibits a characteristic frequency at the rotor frequency and its harmonics (1P, 2P, 3P), evidenced by its PSD in Figure \ref{fig:psd_tip_x_defl}. Turbulence broadens this distribution, raising the mean deflection in the torque-controlled regime. In above-rated conditions, the pitch actuation reduces the aerodynamic loads, as shown by the flattening and the shift of the displacement distribution towards lower deflections. By contrast, the edgewise response $u_y$, shown in Figure \ref{fig:violin_deflections_edgewise} is mostly dominated by inertial and gravitational forces and only marginally linked to aerodynamic load, and it exhibits a symmetric bi-modal distribution with peaks corresponding to the right and left halves of the azimuthal plane. These distributions are biased towards negative edgewise displacement because both the projection of gravitational and aerodynamic loads, influenced by the blade twist, have a stronger negative component, i.e. oriented towards the leading edge. Since the reference frame rotates with the blade pitch (positive towards $y$), the static gravitational load projects periodically onto the edgewise axis, resulting in a strong 1P component shown in Figure \ref{fig:psd_tip_y_defl}. The additional peak observed around 6P likely indicates excitation of a structural mode or resonance of the blade in this direction, possibly arising from dynamic coupling or inherent blade anisotropy.  Turbulence increases the variance of the overall distribution, `smoothing' the two peaks. This effect is particularly noticeable at rated conditions with TI=$15\%$. Arguably, this is due to the activation of the pitch control loop that is activated when the estimated speed exceeds the below-rated limit,  influenced by the stronger turbulent fluctuations. At last, the axial displacement is characterised in Figures \ref{fig:violin_deflections_axial} -  \ref{fig:psd_tip_z_defl}. $u_z(L_b)$ shows a marked negative deflection, which becomes more pronounced at rated conditions. It is worth stressing that the axial displacement is geometrically linked to both deflections because a deflected blade offers a shorter projection along the $z$ axis. In what follows, we refer to this coupling of the displacements as ``geometrical''. This coupling is evidenced by its PSD in Figure \ref{fig:psd_tip_z_defl}, that shows peaks t multiples of the rotor frequency ($n$P) consistent with the flapwise behaviour. Much of the negative tip displacement along $z$ at the highest loads, in \ref{fig:violin_deflections_axial}, is due to the blade out of plane deflection. However, in lower deflections, both at low wind speed and above-rated conditions, the inertial and centrifugal effect results in a positive displacement due to the blade elongation. 

\subsection{Proper Orthogonal Decomposition of Blade Vibrations}\label{res_sec_2}

The POD modes of the blade vibrations are computed from an ensemble dataset containing $T_D=8$ full periods of rotations sampled for each unique combination of wind speed, turbulence level and random seed. The data is first mean-centred to focus on the fluctuations from the steady state response. 

\begin{figure}
         \centering
         \includegraphics[width=0.65\textwidth]{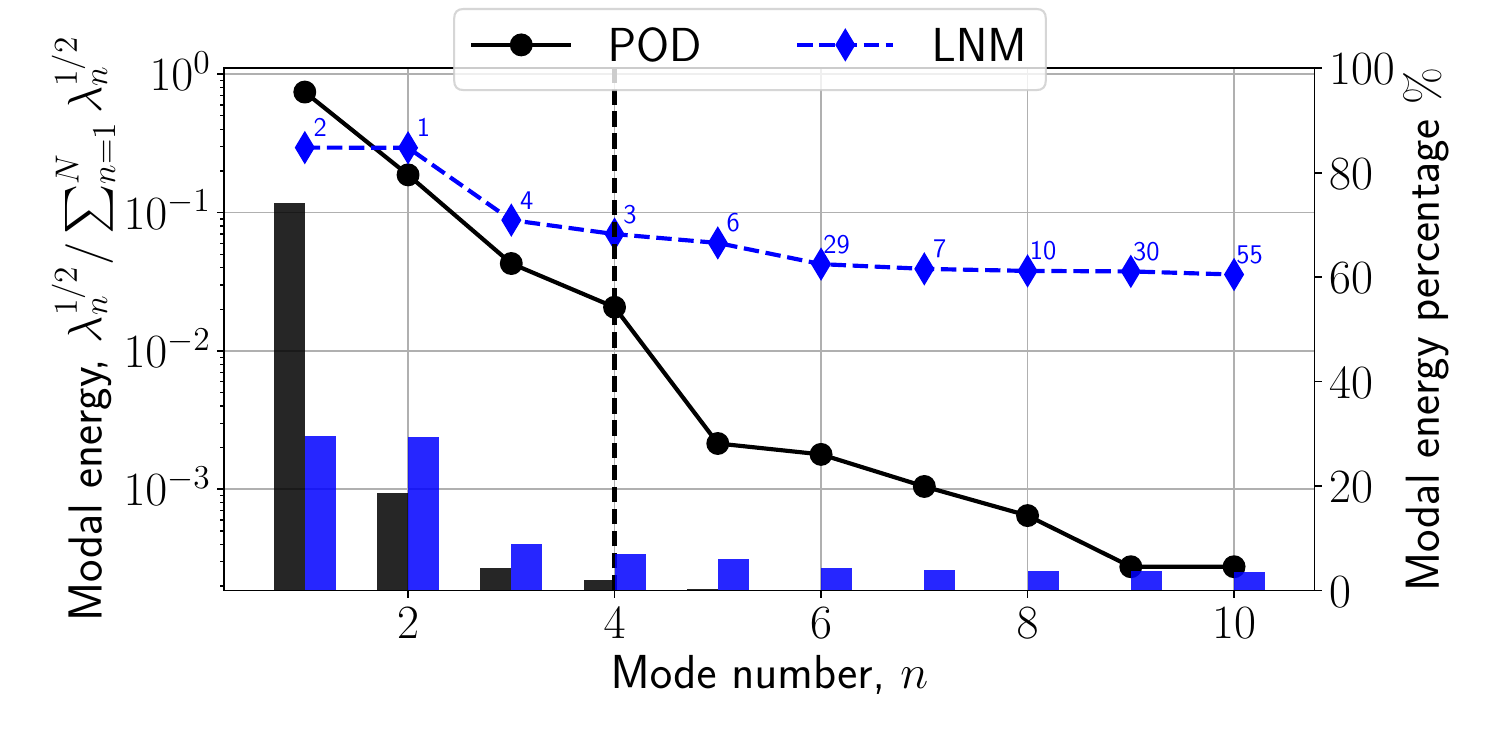}
        \caption{Modal energetic decay of POMs and LNMs. The LNM have been sorted according to their amplitude and \textit{not} according to the associated eigenfrequency $\hat{\omega}_n$, to ease comparison with the POD. The canonical ordering according to $\hat{\omega}_n$ is kept as a reference and indicated on top of each marker. The left $y$ axis shows the energy associated with each mode in the logarithmic axis. The right $y$ axis expresses it in percentage terms computed over the whole modal basis, illustrated by bars.}
        \label{fig:sigmas_tot}
\end{figure}

Figure \ref{fig:sigmas_tot} compares the amplitude of the POD modes against the amplitudes of the LNMs. For recall, LNMs are computed from the generalized eigenvalue problem \citep{gozcu2020representation}

\begin{equation}
    \label{eq:LNM_comp}
    \mathbf{K}\hat{\bm{\phi}}_n = \hat{\omega}_n^2 \mathbf{M} \hat{\bm{\phi}}_n \;,
\end{equation} where $\mathbf{K}$ and $\mathbf{M}$ are the stiffness and mass matrix respectively, $\hat{\bm{\phi}}_n$ are the spatial structures of the LNMs and $\hat{\omega}_n$ the associated (natural) frequencies. LNMs have a harmonic evolution in time; that is, the linear expansion in Eq. \eqref{eq:u_disp_modal_generic} for LNMs gives temporal structures defined as $\hat{a}_n(t)=\exp(\hat{\omega}_n t)$. In this work, we solve the eigenvalue problem \eqref{eq:LNM_comp} only for the eigenfrequencies but infer the corresponding spatial structures of the decomposition $\hat{\bm{\phi}}_n$ directly from the data, similar to the POD. This allows for a simpler comparison by providing modal amplitudes and spatial structures defined and normalized with respect to the same inner product.

Computing a linear decomposition given its spatial or temporal structure is essentially a least square problem \citep{Mendez_2023}. This can be set from the snapshot matrix $\mathbf{D}\in\mathbb{R}^{3n_P\times n_T}$ collecting the deformations along the three directions in each column, and the normalized temporal structure matrix $\hat{\mathbf{\Psi}}\in\mathbb{R}^{n_t\times \hat{n}}$ collecting, in each column, the normalized harmonic evolutions $\psi_n(t_k)=\hat{a}_n(t_k)/||\hat{a}_n(t_k)||_2$ sampled on the $(t_k)^{n_t}_{k=1}$ available time steps. Then, the spatial structures can be computed as 

\begin{equation}
    \label{eq:inv_temp_lnm_2}
    \hat{\mathbf{\Phi}} \hat{\mathbf{\Sigma}} = \mathbf{D} \mathbf{\Psi} \big( \hat{\mathbf{\Psi}}^\top \mathbf{\Psi})^{-1} \;,
\end{equation} where the matrix $\hat{\mathbf{\Phi}}\in\mathbb{R}^{3n_s\times \hat{n}}$ collects the spatial structures of the LNMs and the diagonal matrix $\hat{\mathbf{\Sigma}}\in\mathbb{R}^{\hat{n}\times \hat{n}}$ collects the amplitude of the modes such that also the spatial structure have unitary $l_2$ norm like the POD structures. These amplitude can be computed via normalization of the columns of the matrix on the left hand side of Eq. \eqref{eq:inv_temp_lnm_2}.

To ease the comparison between the modal amplitudes of both decompositions in Figure \ref{fig:sigmas_tot}, the LMNs are sorted in decreasing order of amplitude and not in the traditional ordering based on the associated eigenfrequency. For completeness, the ordering based on increasing natural frequency is indicated on the markers for the LNM decomposition.

The amplitude decay of the POD modes is noticeably sharper than that of the LNMs, indicating that fewer POD modes are required to capture a given amount of variance in the data. In contrast, the LNM amplitudes exhibit an almost asymptotic behaviour from $n\geq 6$. For the POD, a pronounced drop occurs at $n\geq 4$, suggesting that a four-mode truncation may offer a good compromise between parsimony and energy retention, and thereby maintain a high level of approximation accuracy. Thus, in what follows we truncate the POD expansion at $n=4$.

\begin{figure}[h!]
         \centering
         \includegraphics[width=0.95\textwidth]{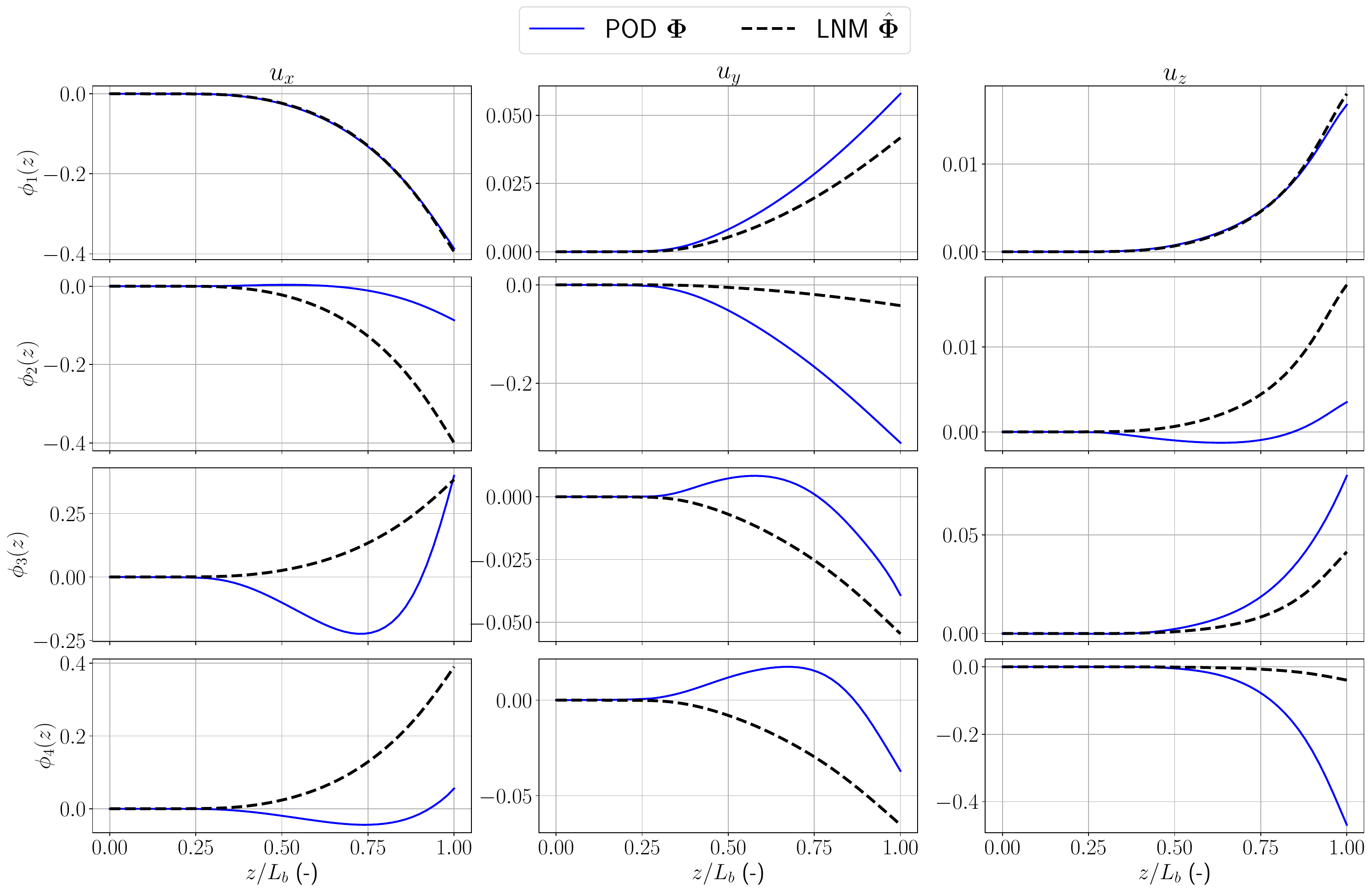}
        \caption{Comparison of POD modes, indicated as solid blue lines, and LNMs expansions computed for the stand-alone blade, depicted with dashed black lines. The modal expansions are truncated at $n=4$ (see Figure \ref{fig:sigmas_tot}).}
        \label{fig:phis}
\end{figure} 

We present the two sets of modes in Figure \ref{fig:phis}. The first POD mode matches the fundamental bending mode, showing that these capture the dominant deformation under operational excitation, while presenting a slightly accentuated edgewise response. Higher POD modes progressively diverge from their LNM counterparts, reflecting the fundamental difference between resonant (described by LNM) and forced response (captured by POD). The second mode, $\bm{\Phi}_2(z)$, is dominated by edgewise motion with a peak near mid-span, suggesting excitation from lateral load components. It also displays localized flapwise tip motion and a twisting-like axial deformation around $z/L_b \approx 0.8$. The third and fourth modes, contributing $\sim4.3\%$ and $\sim2.1\%$ of the variance respectively, capture localized deformations. The third mode shows flapwise bending with a peak at $z/L_b \approx 0.75$ and a sign change in edgewise displacement towards the tip, indicating load redistribution. The fourth mode is characterized by axial bending near the tip, with minor contributions in the flapwise and edgewise directions.
\begin{figure}[h!]
\centering
     \begin{subfigure}[b]{0.44\linewidth}
         % \centering
         \includegraphics[width=\linewidth]{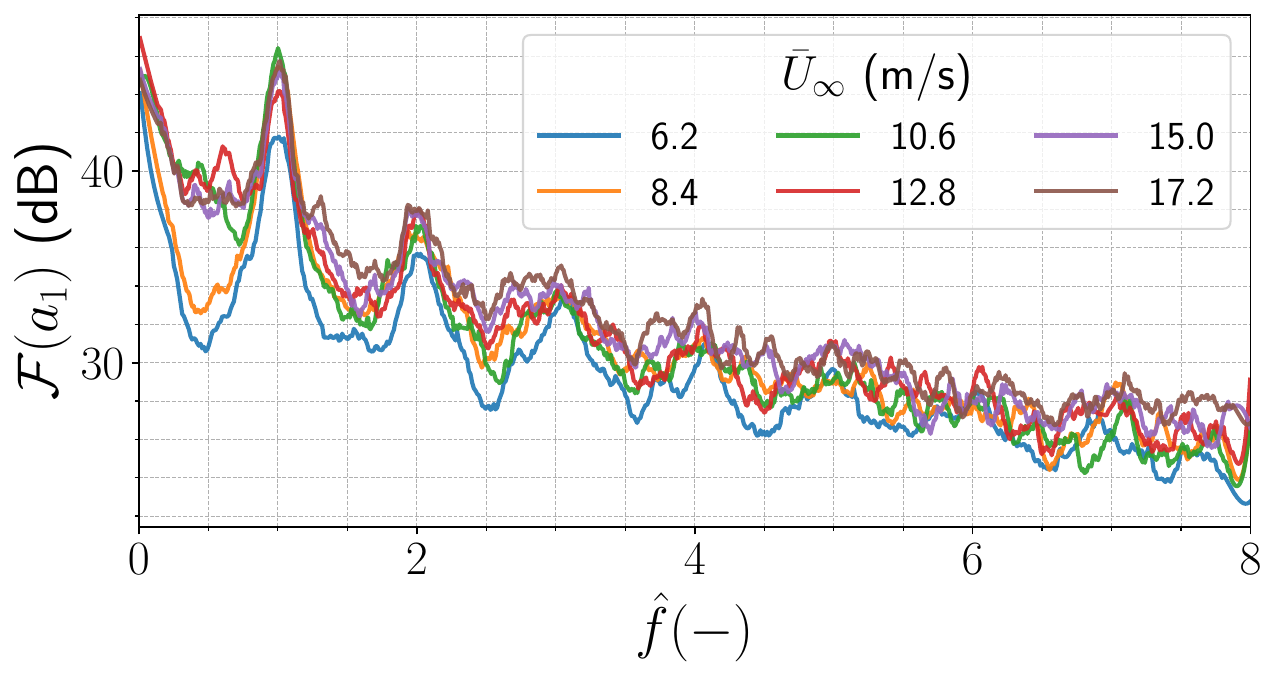}
         \caption{$\mathcal{F}\Big(a_1(t)\Big)$}
         \label{fig:a_r_1_f}
     \end{subfigure}
     \hfill
     \begin{subfigure}[b]{0.44\linewidth}
         % \centering
         \includegraphics[width=\linewidth]{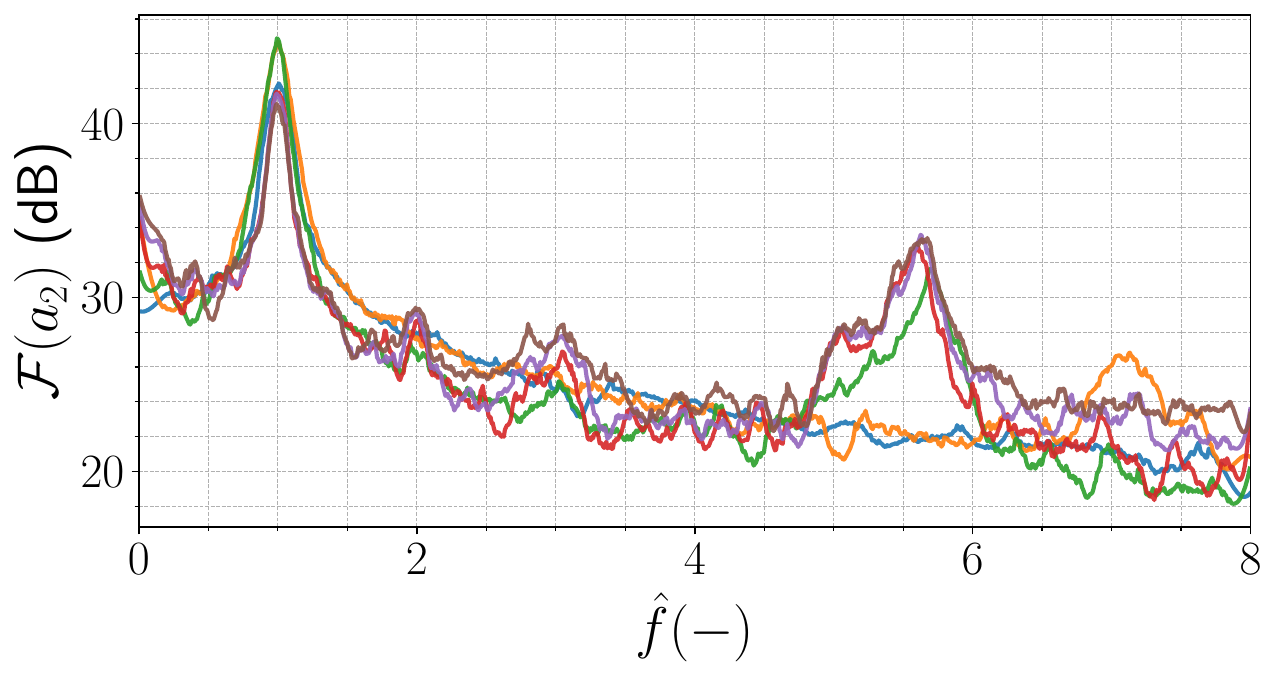}
         \caption{$\mathcal{F}\Big(a_2(t)\Big)$}
         \label{fig:a_r_2_f}
     \end{subfigure}
    \begin{subfigure}[b]{0.44\linewidth}
         % \centering
         \includegraphics[width=\linewidth]{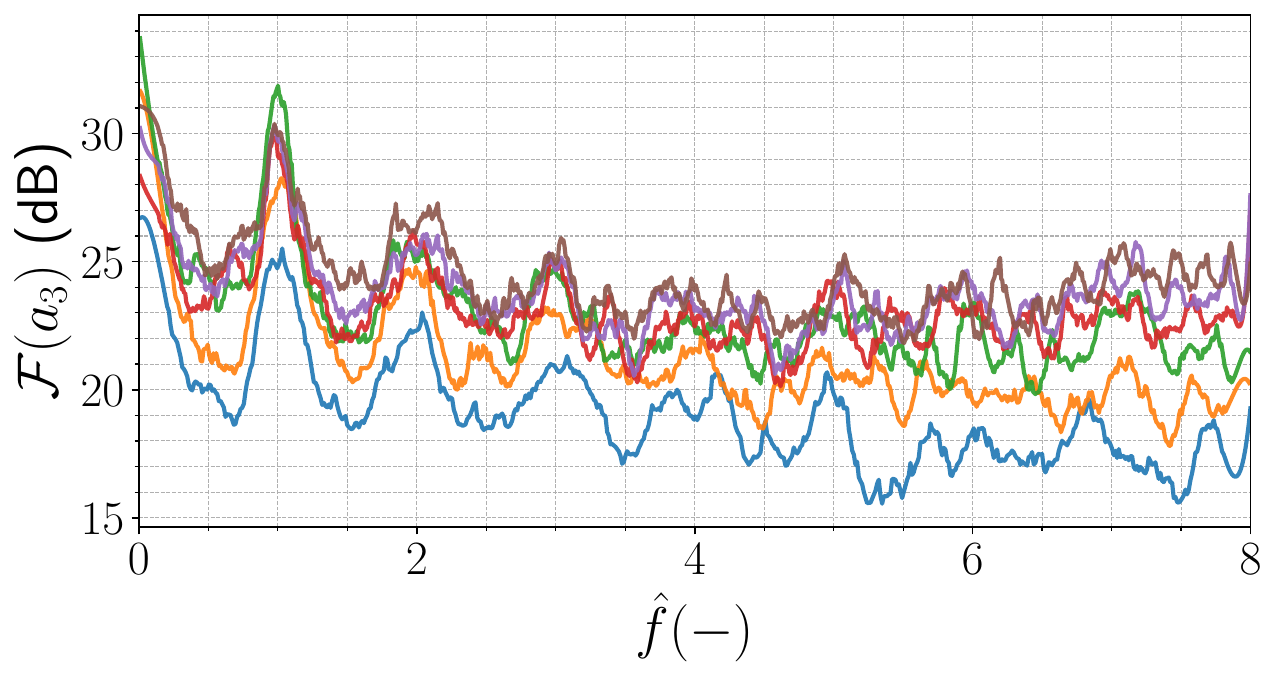}
         \caption{$\mathcal{F}\Big(a_3(t)\Big)$}
         \label{fig:a_r_3_f}
     \end{subfigure}
     \hfill
     \begin{subfigure}[b]{0.44\linewidth}
         % \centering
         \includegraphics[width=\linewidth]{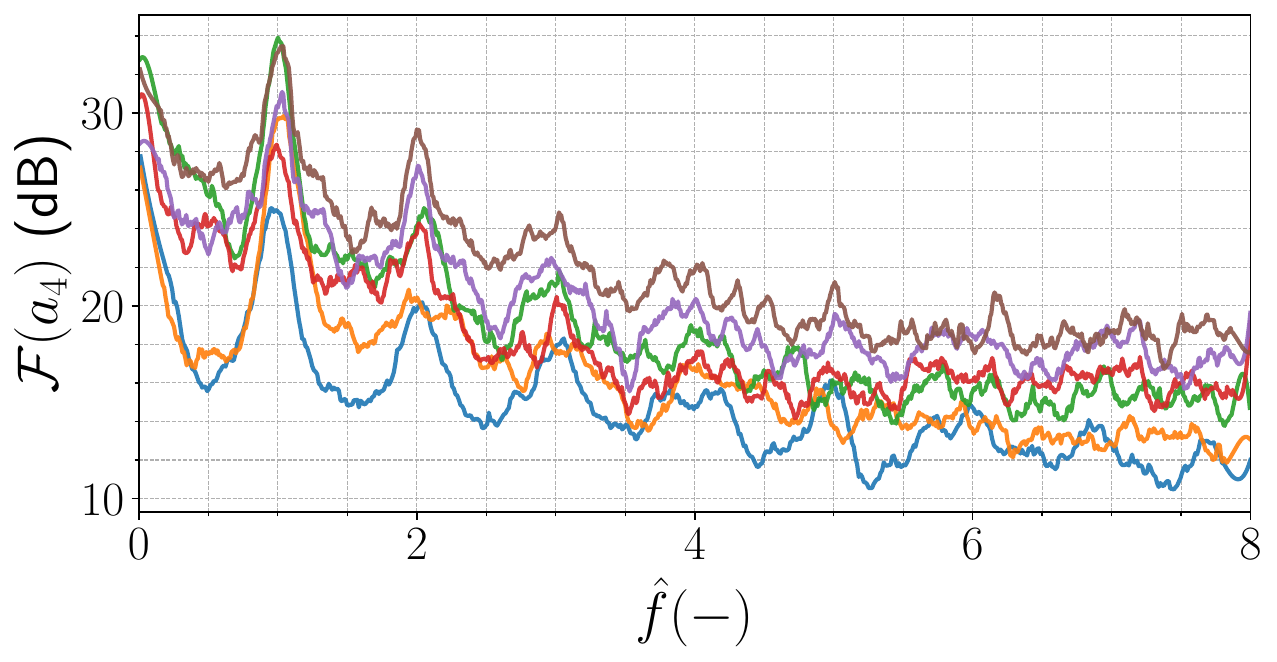}
         \caption{$\mathcal{F}\Big(a_4(t)\Big)$}
         \label{fig:a_r_4_f}
     \end{subfigure}
    \caption{Frequency analysis of the POD modal coefficients for TI=10\% over all wind realisations. Each series is composed by 80000 samples, and it is smoothed via a Savgol filter with a window size of 33 and a third-order polynomial.}
    \label{fig:a_modes_f}
\end{figure} 
The modal dynamics is investigated in Figure \ref{fig:a_modes_f}, collecting the spectra of the POD temporal coefficients $\bm{a}(t)$ (obtained by Eq. \eqref{eq:a_t_modal_proj}). These illustrate how wind speed influences the blade’s dynamic response in the POD space, reflecting the balance of loading sources. These spectra are shown to preserve the frequency signature of the motions in the original space (shown in Figure \ref{fig:tip_defl_u_ti} in the right column), thus being able to retain its coupled nature (presented in more detail in \ref{app_POD_dynamics}). These symmetries between the POD space and the original one highlight the capabilities of the POD to isolate and accentuate distinct dynamical features of the original system. 

\subsection{Azimuthally-Periodic Stochastic ROM of Blade Motions}\label{res_sec_3}
\begin{figure}[h!]
\centering
     \begin{subfigure}[b]{0.8\linewidth}
         % \centering
         \includegraphics[width=\linewidth]{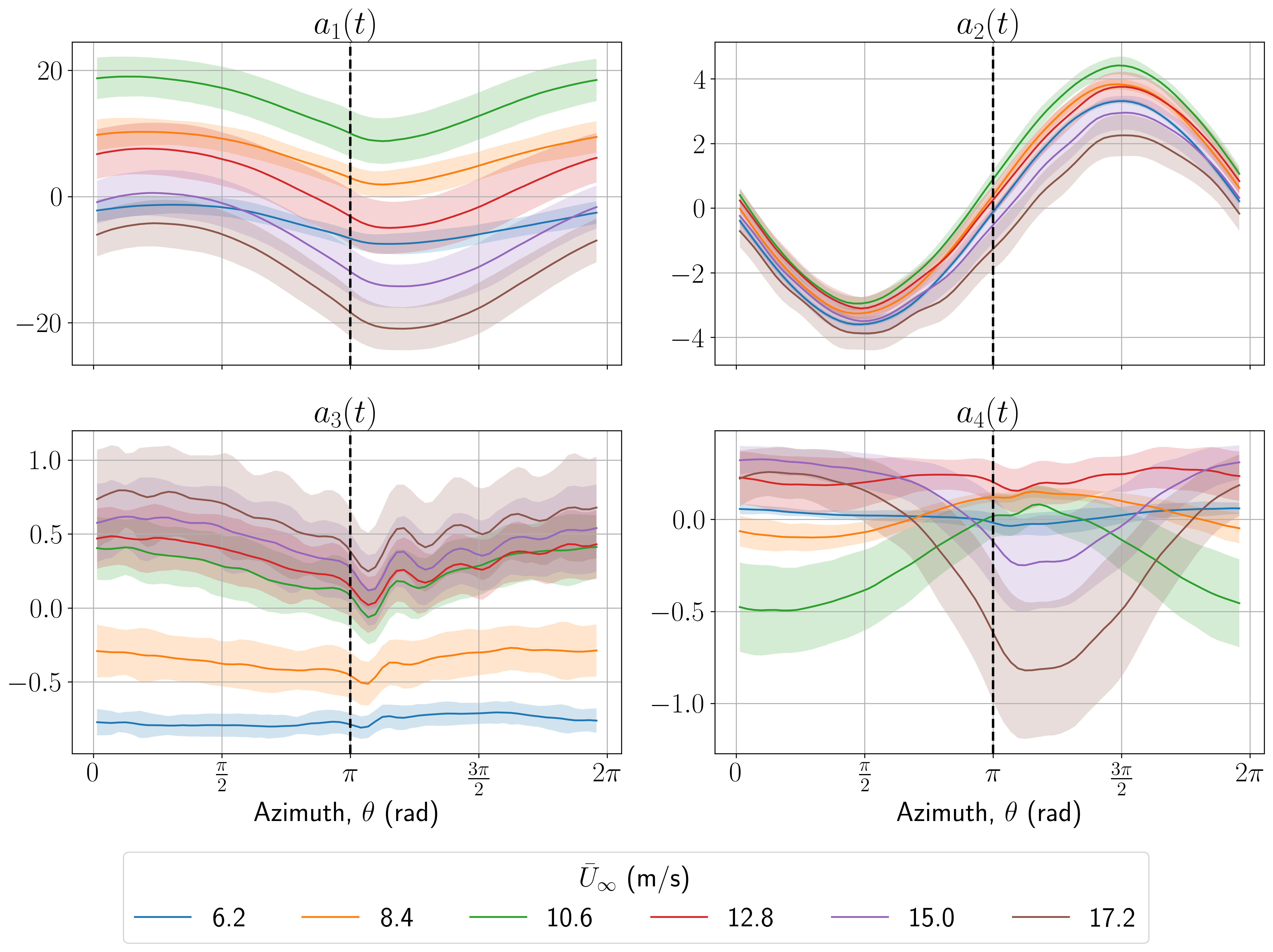}
         \caption{$\bm{a}(\theta; \bar{U}_\infty)$}
         \label{fig:azimuthal_a}
     \end{subfigure}
     \begin{subfigure}[b]{0.8\linewidth}
         % \centering
         \includegraphics[width=\linewidth]{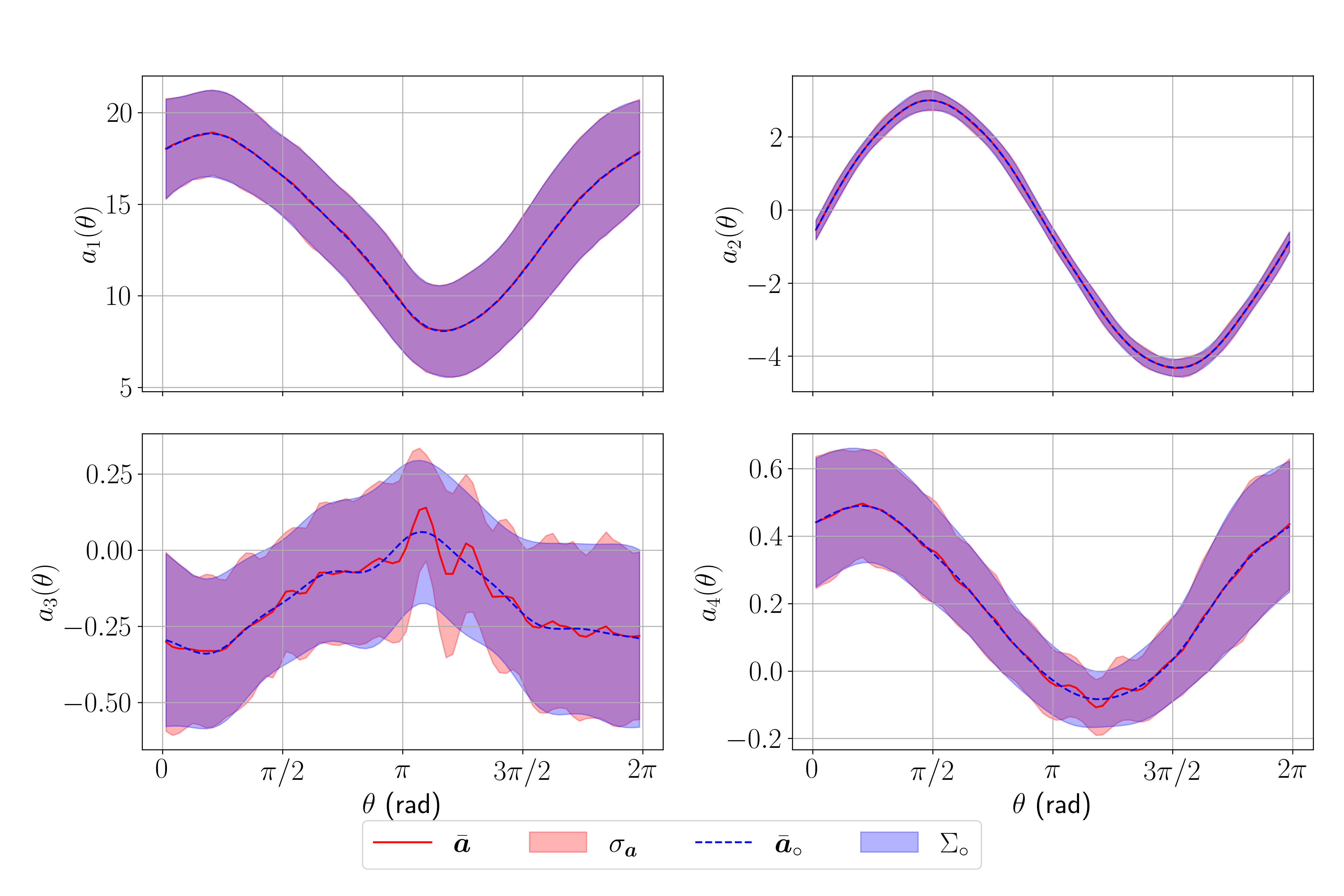}
         \caption{$\bm{a}_\circ(\theta; \bar{U}_\infty)$}
         \label{fig:reconstruction_10_6_Fourier}
     \end{subfigure}
    \caption{Modal dynamics in the azimuthal plane for varying wind conditions (Figure \ref{fig:azimuthal_a}), and its reconstruction via the Fourier stochastic model (Figure \ref{fig:reconstruction_10_6_Fourier}), assessed for $\bar{U}_\infty=10.6$ m/s and TI=10\%.}
\end{figure}

The time-periodic nature of the wind turbine blade forcing motivates the analysis of the reduced-order dynamics in the azimuthal plane.  The analysis of the modal amplitudes and their connection with the operative condition offer a compact way to assess the load signature of the blade, allowing to understand how the position in the rotor plane influences the blade deflection. This analysis is carried out in Figure \ref{fig:azimuthal_a}. For each of the modal coefficients, global ensemble average are shown as solid lines, with the shaded areas spanning plus and minus one standard deviation around the average. These are obtained by binning $\bm{a}(t)$ every ten degrees of azimuth and concatenating, for a given mean wind speed $\bar{U}_\infty$, all random seeds and turbulence conditions. The dashed vertical line at $\theta = \pi$ indicates the tower location.
The first mode ($a_1$) grows in the below-rated regime (blue to green curves) and gradually decreases in the above-rated condition (green to brown) as a consequence of pitch actuation, supporting its correlation with flapwise motions. By contrast, $a_2$ presents an (almost) purely sinusoidal pattern, locally skewed by a combination of rotational and control-induced effects. This confirms its strong physical connection with the inertially-driven edgewise mode hinted in §\ref{res_sec_2}. It exhibits a small magnitude in the upright and downright position of the blade, when gravity is fully aligned, for then increasing as the blade is in the `falling' ($\theta:0\rightarrow\pi$) or `rising' ($\theta: \pi \rightarrow 2\pi$) regime. The azimuthal dynamics of the third mode ($a_3$) shows growing magnitude for increased wind speed, that consistently exhibits a dip around $\theta \approx \pi$ followed by an `overshoot' - indicative of a strong response to tower passage effects. This overshoot triggers an onset of oscillations that persists up to the upright position, where it is eventually damped out. The fourth modes contributes marginally in below-rated conditions and is nearly silent at the lowest wind speed ($\bar{U}_\infty=6.2$ m/s). As wind speed increases, its participation to the overall blade response grows and it presents a sign-inversion when transitioning from below to above rated wind speeds. In below rated, its peak is located near $\theta \approx 2\pi$, for then shifting to $\theta \approx \pi$ in above-rated conditions. Arguably, the former embodies a coupling with the flapwise direction, while the latter captures tower-induced effects.

The Fourier stochastic model (§\ref{sect:qs_model}) is now introduced, to capture both the primary azimuthal dynamical trends and also provides a quantification of their inherent covariance in operational conditions. We illustrate the outcome of this modelling step for $\bar{U}_\infty = 10.6$ m/s and TI=$10\%$ in Figure \ref{fig:reconstruction_10_6_Fourier}, comparing the binned modal trends for this condition, shown by a red solid line and shaded red area for its standard deviation, and the one predicted by the azimuthal Fourier model, illustrated by blue dashed line for the mean and blue shaded area for its deviation. The model well captures the main modal behaviours, while presenting some small mismatches in the mean near the tower passage for the third and fourth POD mode. With this model in place, we proceed now to the full-state deflection field estimation from sparse measurements, for which it provides a foundation for the optimal real-time reconstruction method.

\subsection{Optimal Sensor Placement and Sparse Reconstruction Performances}\label{res_sec_4}

\begin{figure}
    \centering
    \includegraphics[width=0.85\linewidth]{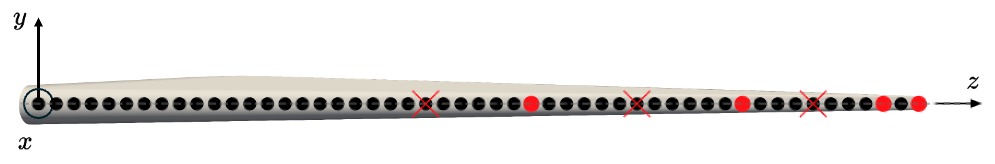}
    \caption{Radial locations of interest. Red markers depict the optimal sensors placement $\mathbf{z}_p=(1.0, 0.8, 0.56, 0.96)L_b$, sorted in decreasing order of importance. The red crosses illustrate the points chosen to qualitatively examine the reconstruction in unmeasured locations $\mathbf{z}_o=(0.44, 0.68, 0.88)L_b$.}
    \label{fig:osp}
\end{figure}

The ability to reconstruct high-dimensional fields from sparse data critically relies on the sensor locations and the physics they capture. To determine the positions that maximise the modal sensing efficiency, we used the pivoted QR decomposition applied to the leading POD modes (§\ref{sect:sparse_methods}). We show the result of this procedure in Figure \ref{fig:osp}, in which we indicate the resulting locations with circular red markers. 
 The optimal locations, sorted by importance, are $\mathbf{z}_p=(1.0, 0.8, 0.56, 0.96)L_b$. The QR approach ensures that these locations maximise the degree of independence between the retrieved measures and are thus as informative as possible for a linear estimator. That is, these locations provide key insight to infer the modal shapes $\bm{\Phi}(z)$. To assess the quality of the reconstruction in unmeasured locations, we select the observation points $\mathbf{z}_o$, illustrated by red crosses in Figure \ref{fig:osp}. These locations are chosen from different structural regions to capture different dynamic features and a varying degree of modal superposition. We note that at this stage the POD-based estimation yields fluctuations over the mean deflection field $\bar{\bm{u}}(z)$, removed prior to the decomposition step on the ensemble of realisations of $\bm{u}(z,t)$ and added back before comparing with the true fields. Moreover, in what follows we consider a measurement standard deviation $\sigma_\bullet^{(p)}=0.1$, that completes the sensing process defined in Equation \eqref{eq:multivariate_meas}. 

\begin{figure}[h!]
\centering
     \begin{subfigure}[b]{0.85\linewidth}
         % \centering
        \includegraphics[width=\linewidth]{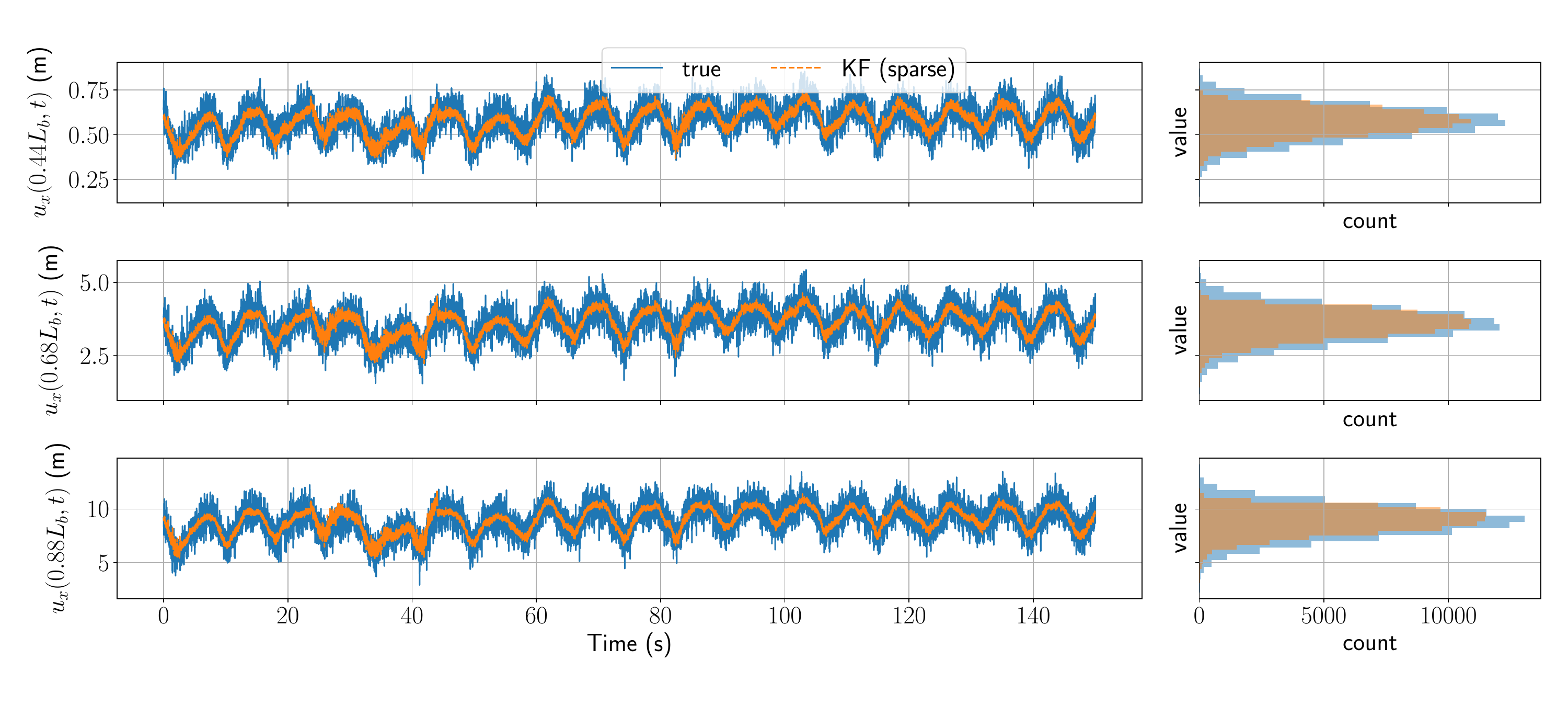}
        \caption{Reconstruction of flapwise motions at $z=\mathbf{z}_o$ for $\bar{U}_\infty=10.6$ m/s.}
        \label{fig:reconstr-x}
     \end{subfigure}
     
    \begin{subfigure}[b]{0.85\linewidth}
         % \centering
        \includegraphics[width=\linewidth]{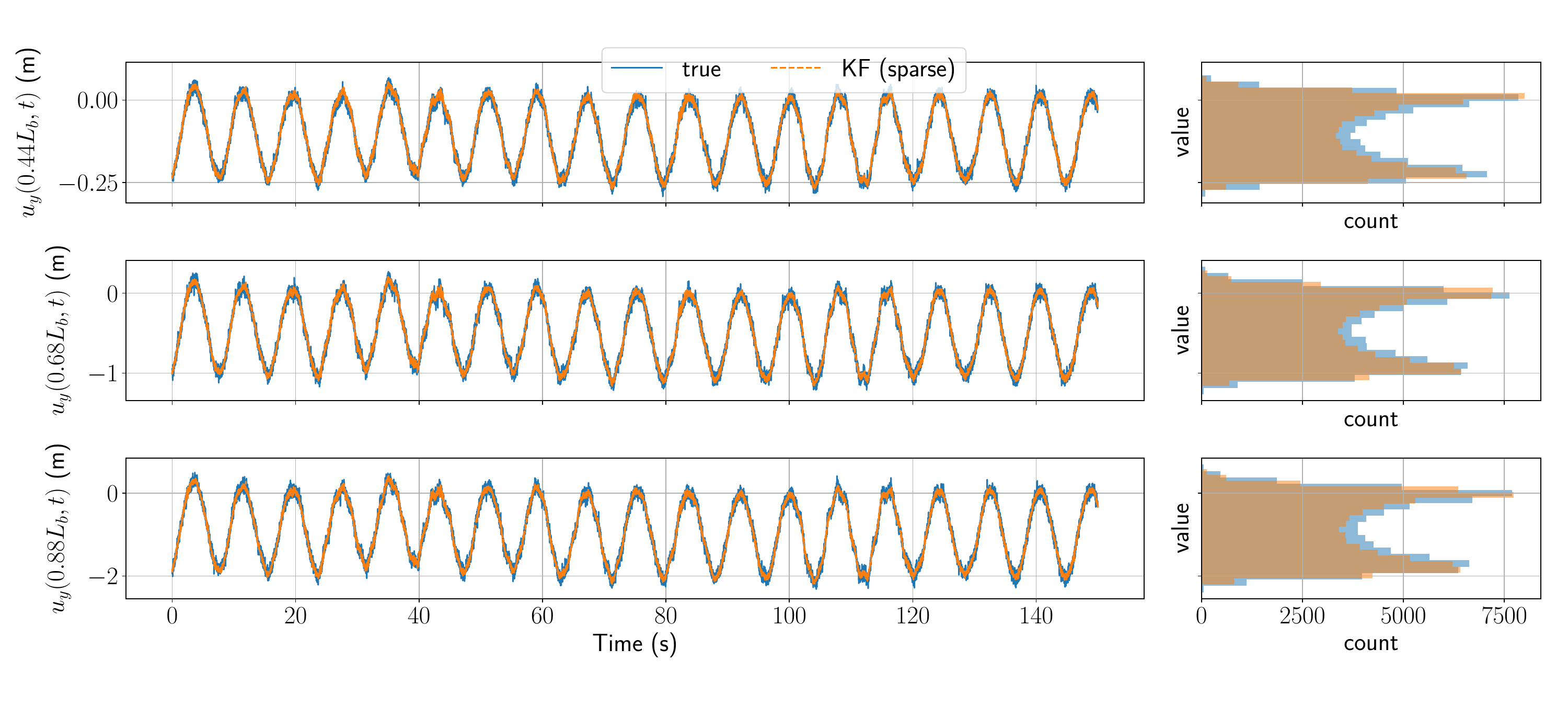}
        \caption{Reconstruction of edgewise motions at $z=\mathbf{z}_o$ for $\bar{U}_\infty=10.6$ m/s.}
        \label{fig:reconstr-y}
     \end{subfigure}
     \hfill
     \begin{subfigure}[b]{0.85\linewidth}
         % \centering
    \includegraphics[width=\linewidth]{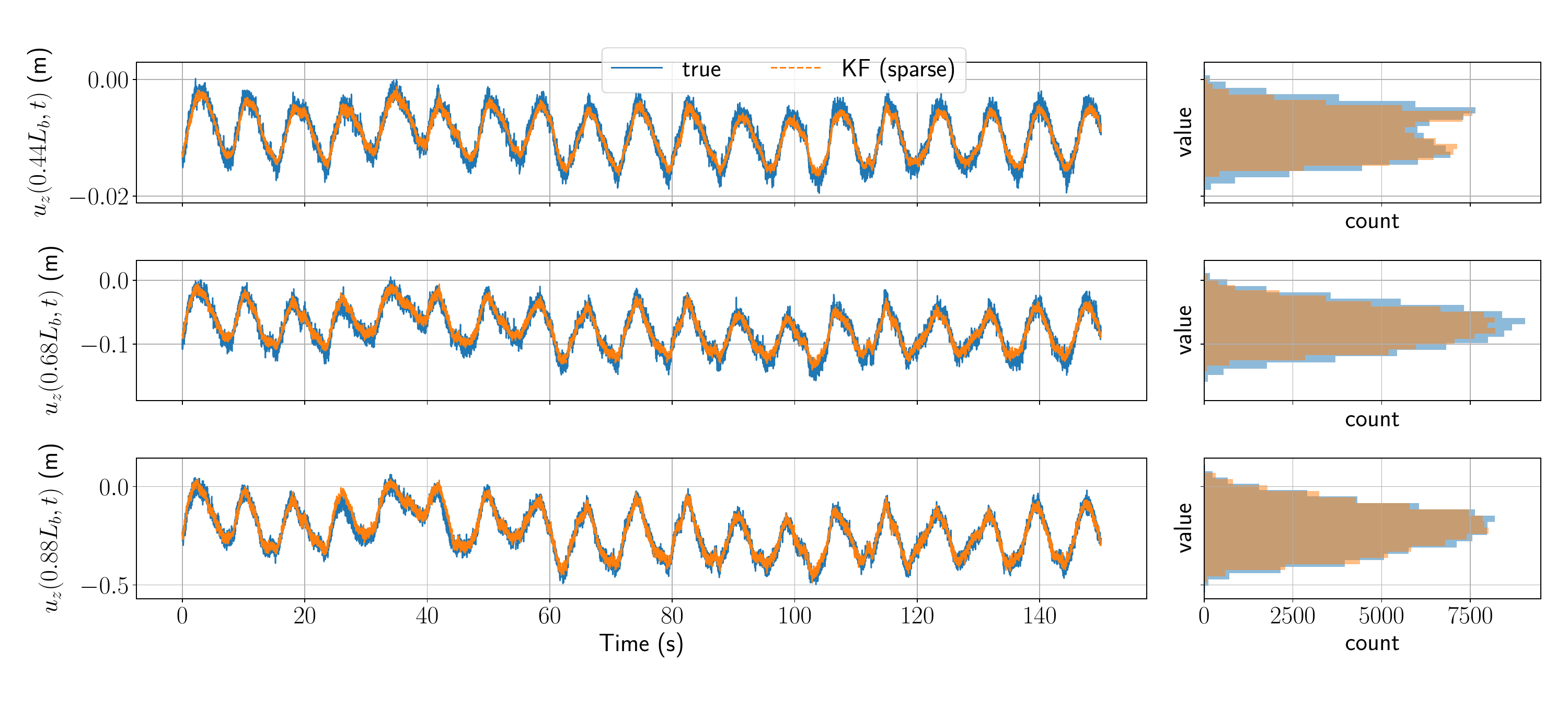}
    \caption{Reconstruction of axial motions at $z=\mathbf{z}_o$ for $\bar{U}_\infty=10.6$ m/s.}
    \label{fig:reconstr-z}
     \end{subfigure}
    \caption{POD-Kalman estimator performances, for the three directions of deflections.}
    \label{fig:podkf}
\end{figure}

We analyse here the estimator performances at peak-loading and most dynamic conditions. Namely, rated operations $\bar{U}_\infty=10.6$ m/s and with a TI=$15\%$. The results of the estimation are shown in Figures \ref{fig:reconstr-x}, \ref{fig:reconstr-y} and \ref{fig:reconstr-z} reconstructing the flapwise, edgewise and axial motions, respectively. 

% \begin{figure}
%     \centering
%     \includegraphics[width=0.9\linewidth]{FIGURES_LAST/reconstruction_x_10_6_TI0_10_KF_None_displacement.pdf}
%     \caption{Reconstruction of flapwise motions at $z=\mathbf{z}_o$ for $\bar{U}_\infty=10.6$ m/s.}
%     \label{fig:reconstr-x}
% \end{figure} 
The reconstruction of the flapwise motions (Figure \ref{fig:reconstr-x}, orange line) closely follows the true, noisy deflections (blue solid lines), capturing both low-frequency and high-frequency deflections with good agreement. The histograms on the right column confirm that the statistical distribution of the reconstructed deflections aligns with the true values with the peaks of such distributions located at approximatively the same values. It can also be noticed that the estimated fields exhibit a narrower distribution with respect to the noisy full-field signal. 
% \begin{figure}
%     \centering
%     \includegraphics[width=0.9\linewidth]{FIGURES_LAST/reconstruction_y_10_6_TI0_10_KF_None_displacement.pdf}
%     \caption{Reconstruction of edgewise motions at $z=\mathbf{z}_o$ for $\bar{U}_\infty=10.6$ m/s.}
%     \label{fig:reconstr-y}
% \end{figure} 
A similar behaviour is observed in analysing the deflections in the edgewise direction, shown in Figure \ref{fig:reconstr-y}. The reconstruction aligns well with the true signals, presenting minimal deviation in both amplitude and phase. This is further confirmed by the histograms, which suggest that the relevant dynamics are successfully reconstructed. 
% \begin{figure}
%     \centering
%     \includegraphics[width=0.9\linewidth]{FIGURES_LAST/reconstruction_z_10_6_TI0_10_KF_None_displacement.pdf}
%     \caption{Reconstruction of axial motions at $z=\mathbf{z}_o$ for $\bar{U}_\infty=10.6$ m/s.}
%     \label{fig:reconstr-z}
% \end{figure} 
Compared to flapwise and edgewise motions, the axial dynamic is characterised by displacements that are one order of magnitude smaller, as illustrated in Figure \ref{fig:reconstr-z}. In the axial direction, the filtering action of the estimator is particularly beneficial, as noise and other effects could otherwise obscure the physically relevant response. 

\begin{figure}
    \centering
    \includegraphics[width=0.8\linewidth]{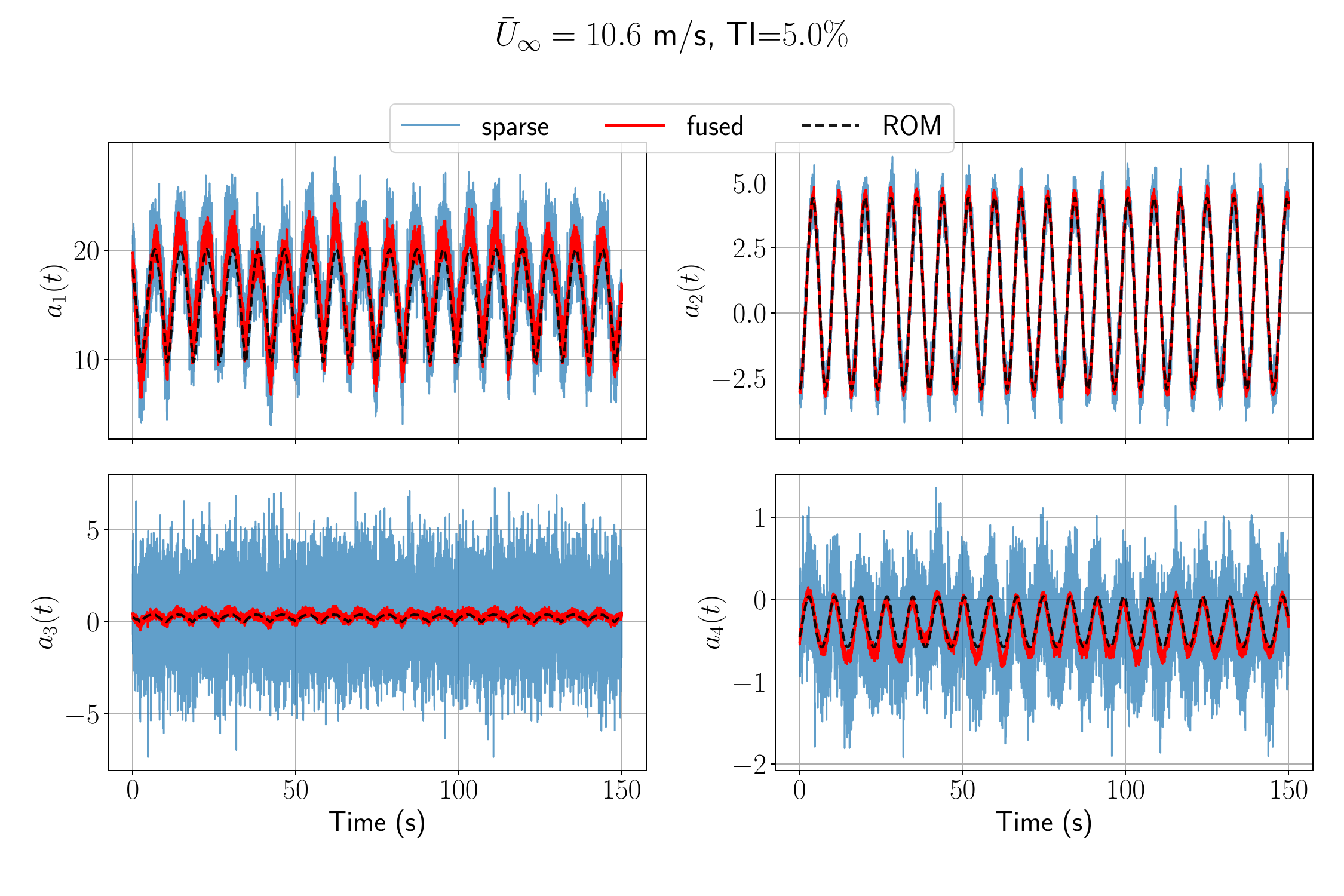}
    \caption{Individual contributions to final estimation. `Sparse' indicates direct projection from sensor measurements, `fused' is the combined result of the estimation, and `ROM' is the stochastic model.}
    \label{fig:contributions_KF}
\end{figure}

These results demonstrate the effectiveness of the sparse Kalman reconstruction in accurately capturing the dominant three-dimensional blade motions under dynamic conditions. The estimator combines the \textit{a priori} filtering due to the POD basis, and the \textit{a posteriori} Kalman corrections, leading to robust measures. It is worth to investigate the relative weight of these individual contributions of the information sources (model, measures) to the overall estimated field. Given that this estimator relies on a modal decomposition, it is natural to carry out this comparison in the modal space, in which the contributions of each mode to the overall dynamics can be assessed separately. This inspection is shown in Figure \ref{fig:contributions_KF}, for the same wind speed of the previous reconstruction but reducing the turbulence intensity to TI=$5\%$ to ease the visualisation.
 Figure \ref{fig:contributions_KF} shows that the model effectively acts as a stabilising prior, leading to a physically meaningful reconstruction. This demonstrates the effectiveness of Kalman-based approaches in prioritising the most reliable source of information, e.g. the one characterised by lower variance. This balance appears to play a major role in the context of this work, in which leading modes benefit from measurement corrections and higher-order modes, whose direct estimation would be almost undistinguishable from noise, are inferred by the physics-based ROM. Moreover, the robust tracking of the dynamics in the POD subspace opens the path for the joint inference of unmeasured quantities, of which we give an example in \ref{app_1}.

\section{Conclusions and Outlook}\label{sect:conclusions}

This study presents a Kalman-based, real-time, estimator of full blade deformations from sparse measurements. First, in the calibration phase, we use dense sensor configuration to assemble a dataset containing the three-dimensional blade dynamics corresponding to varying operational conditions. We then compute the Proper Orthogonal Decomposition (POD) of blade motions, reducing the high-dimensional dynamics to a subspace spanned by an essential set of modes retaining most of the physics associated with the operational forced response. The leading POD mode exhibits a strong connection the fundamental response of the blades, while higher order modes capture localised dynamics that substantially differ from the LNMs. 

This reduced-order representation is found to ease the description of the dynamical traits of the blade, effectively separating the features of its behaviour while preserving their associated physics. We then formulate an azimuthally-periodic stochastic ROM in the reduced POD space, leveraging the azimuthal periodicity of blade forcing, which provides a low-dimensional prior of the blade's response as it travels across the rotor disk, retaining first and second-order statistics. This constructed model is stationary, and describes dominant pattern of vibration and their associated variability in a given operative condition, defined by hub-height wind speed and turbulence intensity. This provides insights into the dynamics of the blade and its `loads signature' as determined by both external forcing, e.g. atmospheric conditions, and operative decisions, e.g. control actions. 

The POD basis also guides the measurement process, as it is used to determine the strategic locations of the limited number of sensors available, to maximise the independence of the retrieved modal information. We assume direct observability of the deflection at the selected measurement points, albeit with noise contamination (corresponding, for instance, to multi-directional strain measurements). The resulting sensor locations are effective in capturing both large scale and localised effects, yielding an accurate modal reconstruction with minimal instrumentation.

At last, the measurements and the ROM are integrated by a Kalman-based fusion operation that optimally balances measurement-driven reconstruction with the azimutal model regularisation. The results show that this filtering process is beneficial for the estimation of the dynamics in the reduced-order space, particularly for higher-order modal coordinates that appear more sensitive to undersampling and noise. Notably, the method achieves accurate reconstruction of the distribution of blade deflections across the whole radial span, maintaining consistency across all operative conditions tested. Notably, the robust identification of the dynamics in the POD subspace opens the path to the inference of unmeasured quantities, and we illustrate this application reconstructing the blade torsion in \ref{app_1}.

In conclusion, the proposed estimator accurately reconstructs blade motions from sparse and noisy measurements. This enables real-time monitoring of wind turbine structural response and may be extended to other dynamic features, such as tower motion. We believe this framework offers a path to the quantitative assessment of the influence of atmospheric factors (e.g., gusts) and operational conditions (e.g., control laws) on turbine loading while being broadly applicable to other engineering systems where direct sensing of the full-dimensional fields is impractical.

\section*{Acknowledgements}

L. Schena is supported by Fonds Wetenschappelijk Onderzoek (FWO), Proj. Number 1S67925N. This project has been partially funded by the European Research Council (ERC, grant agreement No 101165479 RE-TWIST StG). Views and opinions expressed are however those of the authors only and do not necessarily reflect those of the European Union or the European Research Council. Neither the European Union nor the granting authority can be held responsible for them.

\clearpage

\appendix
\section{Reduced-Order Representation of Wind Turbine Blade's Dynamical traits}\label{app_POD_dynamics}

\begin{figure}
\centering
     \begin{subfigure}[b]{0.9\linewidth}
         % \centering
         \includegraphics[width=\linewidth]{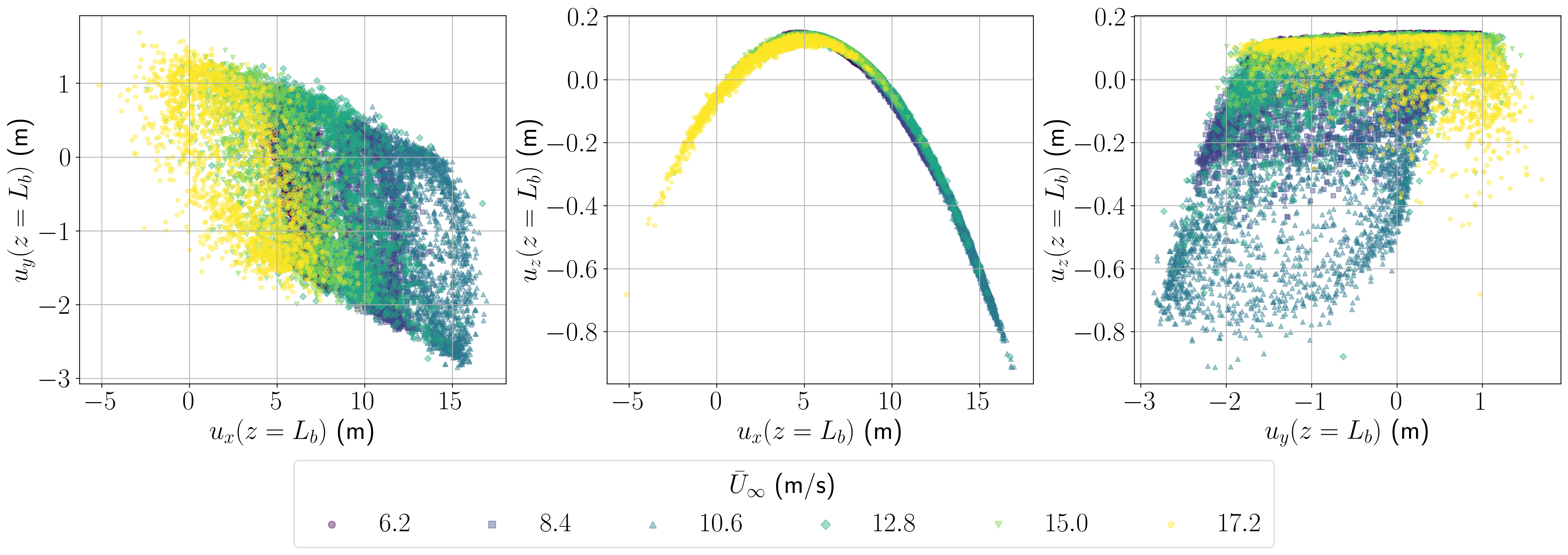}
         \caption{Blade tip couplings in the original space.}
         \label{fig:blade_tip_coupling_full}
     \end{subfigure}

     \begin{subfigure}[b]{0.9\linewidth}
         % \centering
         \includegraphics[width=\linewidth]{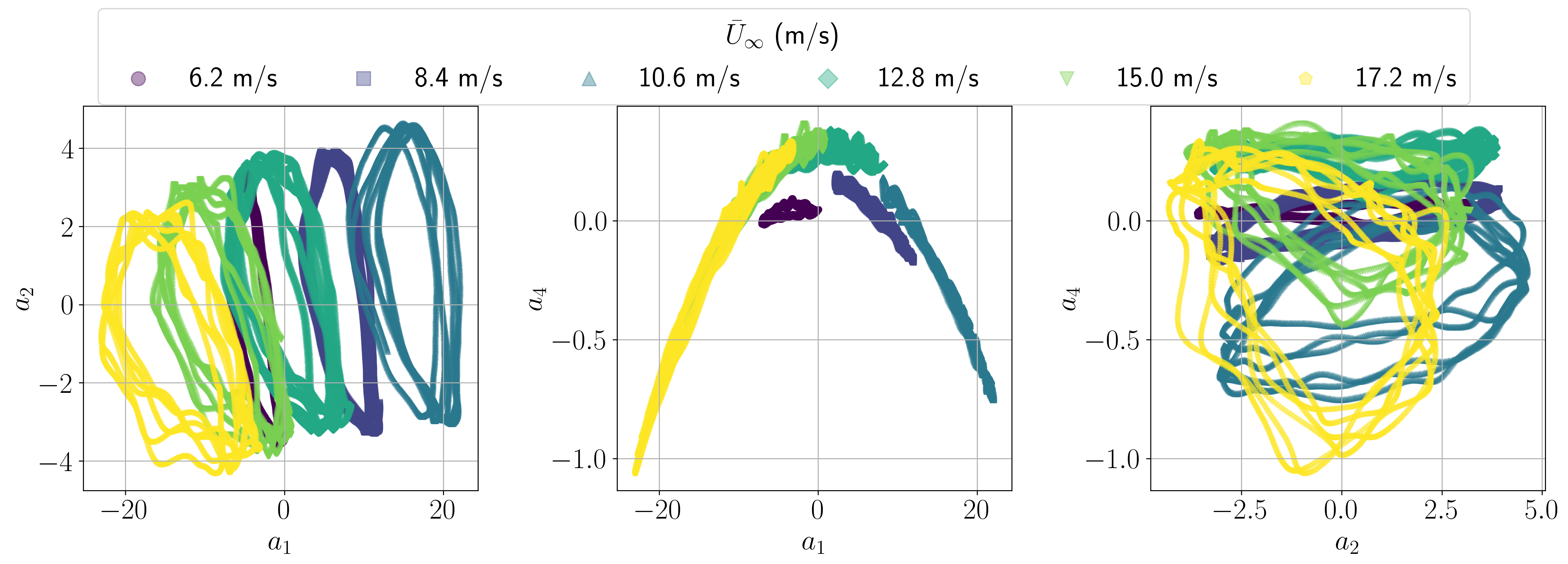}
         \caption{Couplings between POD modal coefficients.}
         \label{fig:pod_coupling}
     \end{subfigure}
    \caption{Comparison of the full-space and POD space coupling, shown for different wind speeds.}
    \label{fig:couplings}
\end{figure}

Figure \ref{fig:couplings}, that examines coupled effects between different displacement directions (Figure \ref{fig:blade_tip_coupling_full}) and their modal counterpart (Figure \ref{fig:pod_coupling}). The dynamics in the POD space effectively capture the driving patterns in the high-dimensional space, namely: (1) the out-of-phase relationship between the flapwise and edgewise modes $u_x-u_y$ is preserved in the $a_1-a_2$ coupling the gradual change of orientation observed in the tip responses; (2) the geometric coupling between $u_x-u_z$ is also observed in the $a_1-a_4$ plane; and (3) the elliptical trajectories between $u_y-u_z$ are characteristic of the $a_2-a_4$ response. 

\section{POD-based inference of unmeasured quantities: a torsional example}\label{app_1}
%Yet, blade dynamics should not be intended as a rigid-body motion characterised by deflections alone.  On the opposite, it appears to be characterised by sectional rotations, more pronounced in modern large designs, that while being intimately connected to $\bm{u}(z, t)$ describes a substantially different dynamics. Their combined effect greatly influences not only the blade structural response \citep{jeong2014effects, jensen2023torsional}, but also the flow characteristics downwind the machine \citep{trigaux2024investigation}, potentially affecting subsequent machines. 
The previous discussion carried out in this article focused on the reconstruction of blade deflections, $\bm{u}(z,t)$ (§\ref{res_sec_1}), and its compact representation in a reduced-order space spanned by the POD modes described by the modal coefficients $\bm{a}(t)$ (§\ref{res_sec_2}). It is natural to ask whether we could leverage the estimated deflection fields to retrieve information about the unmeasured blade torsional state, a standard task in the context of POD-based Linear Stochastic Estimation \citep{adrian1988stochastic, bonnet1994stochastic, podvin2018combining}. To this end, we define the sectional rotations\footnote{Expressed as Wiener-Milenkovic parameters, see \citep{wang2017beamdyn}.} in the three directions as $\bm{\tau}(z, t) = (\tau_x(z,t), \tau_y(z,t), \tau_z(z,t))$, in which the first two indicate twisting in the relative directions, and the last embodies pure torsion on the $z$ axis. The generic modal representation of this field (analogous to Equation \eqref{eq:u_disp_modal_generic}) reads
\begin{equation}
    \label{eq:tors_modal_generic}
    \bm{\tau}(z,t) \approx \sum_{j=1}^J \bm{\xi}_j(z) \; b_j(t) \;,
\end{equation} in which $b_n(t) = \langle \bm{\tau}(z,t), \xi(z)\rangle$ describes the dynamics in the modal space. 

Thus, we seek a linear mapping from $\bm{a}(t)$ to $\bm{b}(t)$ such that
\begin{equation}
    \label{eq:linear_mapping}
    \bm{b}(t) \approx \bm{M} \bm{a}(t) \;.
\end{equation} Finding $\bm{M}$ is a classic least square problem, solved by pseudo-inversion of $\bm{a}(t)$. Once again, we shall use the POD on the collected data to retrieve the modal basis $\bm{\xi}(z)$. 
\begin{figure}
    \centering
    \includegraphics[width=0.85\linewidth]{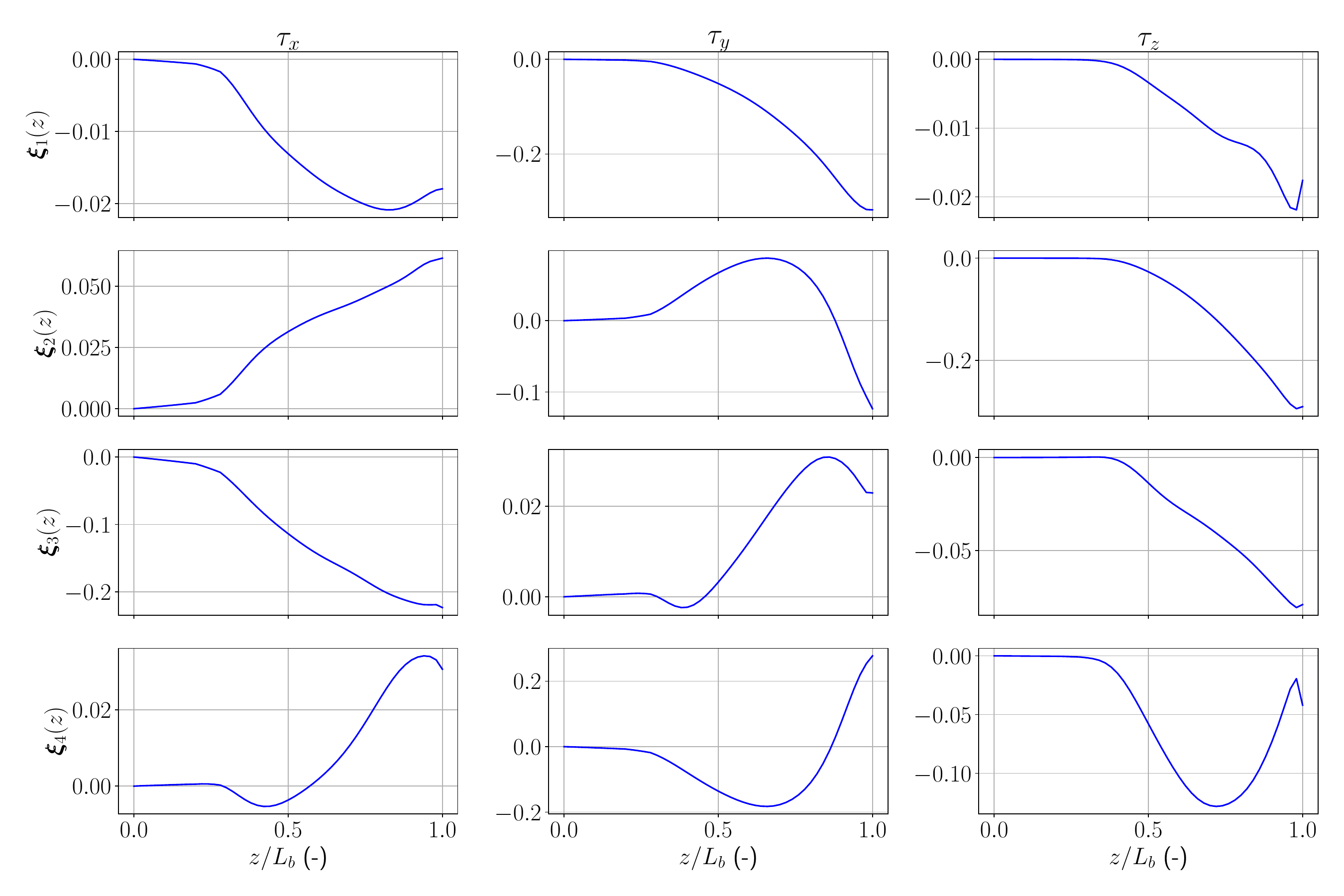}
    \caption{POD modes of torsion, $\bm{\Xi}(z)$.}
    \label{fig:torsional_modes_POD}
\end{figure}

\begin{figure}[h!]
\centering
     \begin{subfigure}[b]{0.32\linewidth}
         % \centering
         \includegraphics[width=\linewidth]{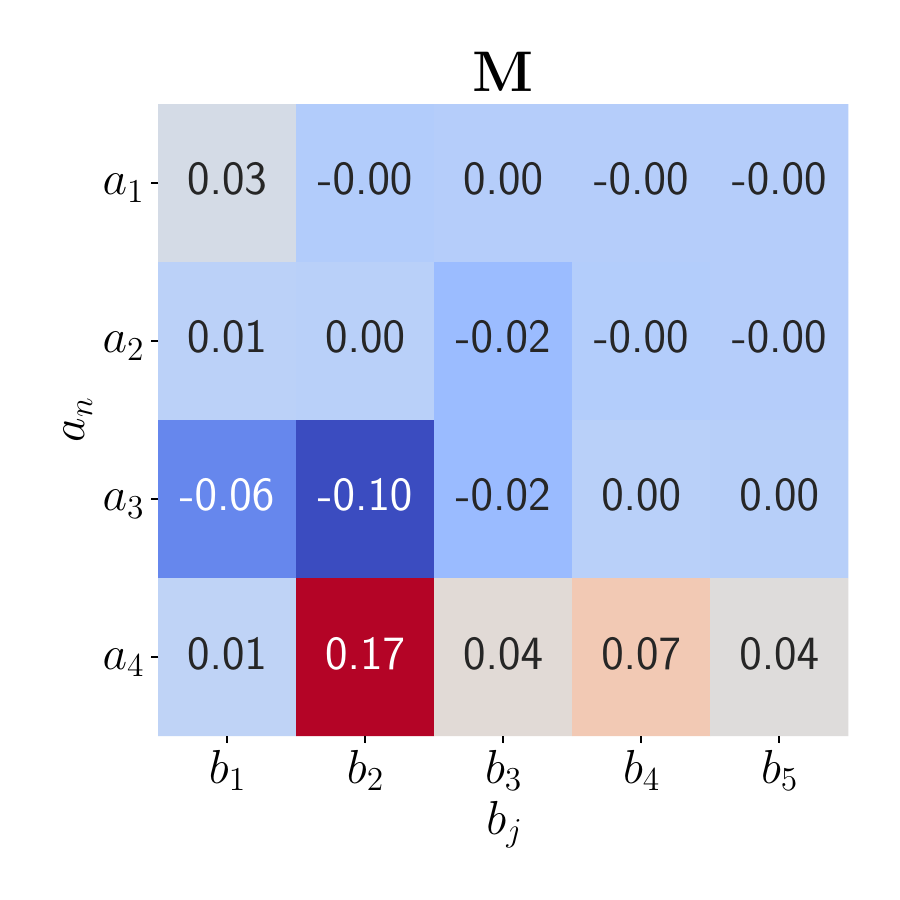}
         \caption{$\bar{U}_\infty =6.2$ m/s.}
         \label{fig:M_6.2}
     \end{subfigure}
     % \hfill
     \begin{subfigure}[b]{0.32\linewidth}
         % \centering
         \includegraphics[width=\linewidth]{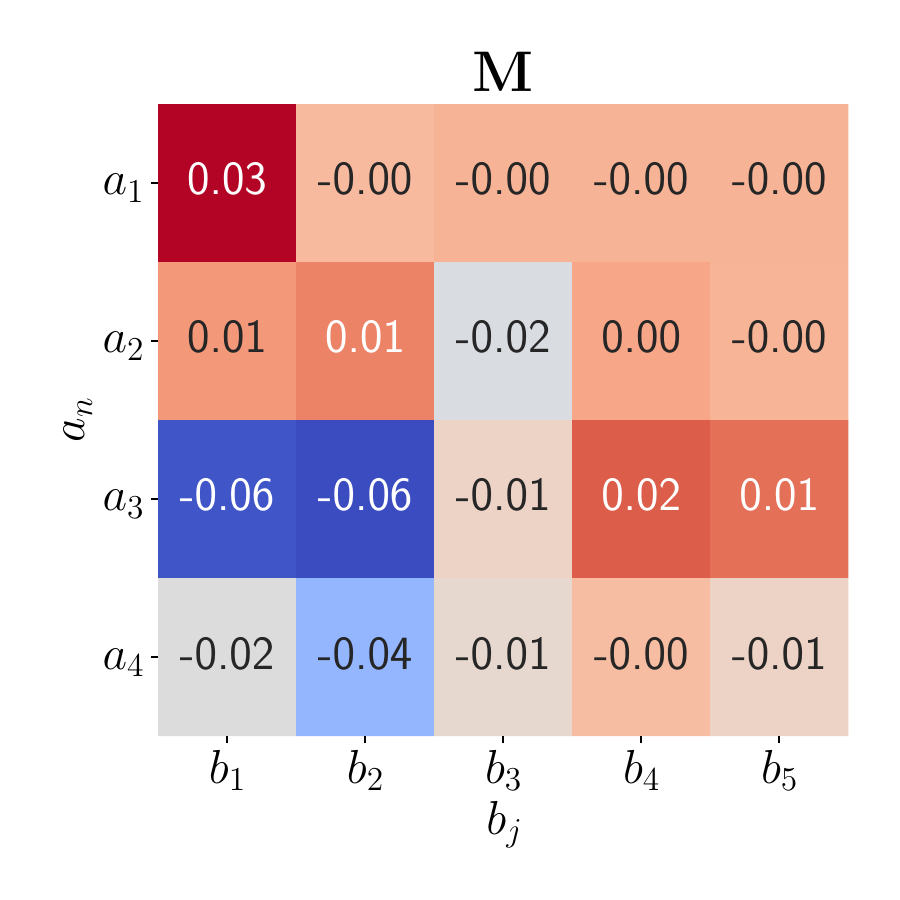}
                  \caption{$\bar{U}_\infty =10.6$ m/s.}
         \label{fig:M_10.6}
     \end{subfigure}
    \begin{subfigure}[b]{0.32\linewidth}
         % \centering
         \includegraphics[width=\linewidth]{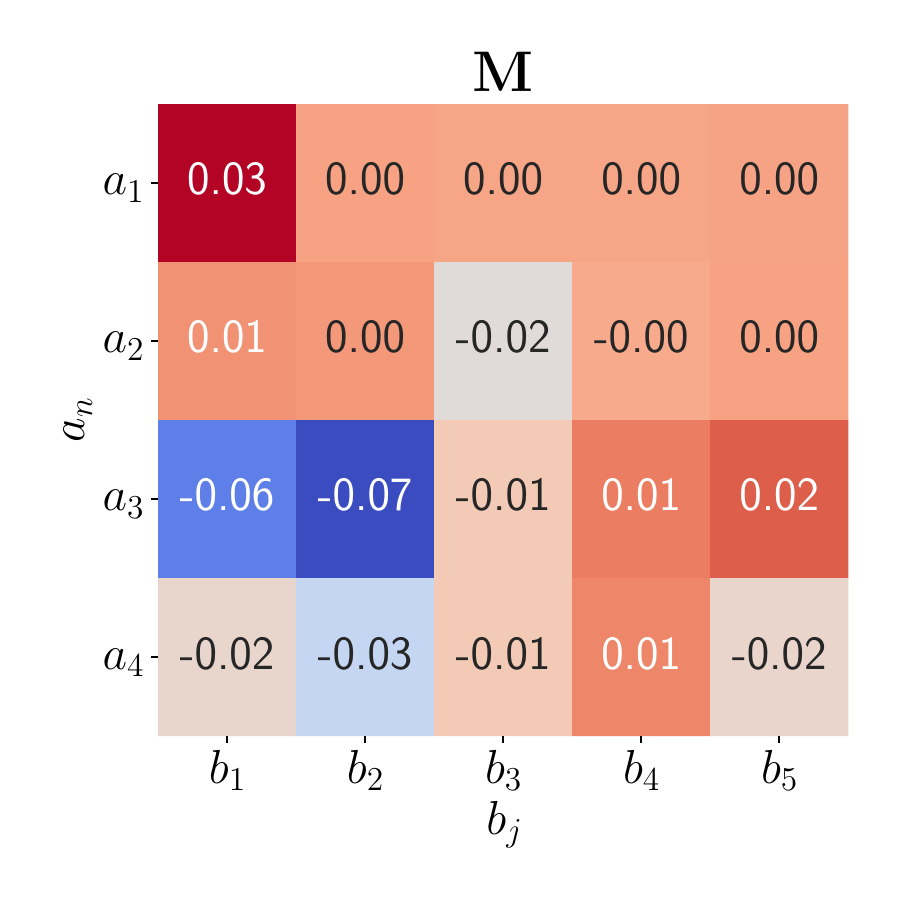}
         \caption{$\bar{U}_\infty =15.0$ m/s.}
         \label{fig:M_15.0}
     \end{subfigure}
    \caption{Variation of $\bm{M}$ with wind speed, evaluated at three different operative points (characterised by TI=$10\%$): below-rated ($\bar{U}_\infty=6.2$ m/s, Figure \ref{fig:M_6.2}), rated ($\bar{U}_\infty=10.6$ m/s, Figure \ref{fig:M_10.6}) and above-rated ($\bar{U}_\infty=15.0$ m/s, Figure \ref{fig:M_15.0}).}
    \label{fig:M_wspd}
\end{figure} The relationship between $\bm{a}(t)$ and $\bm{b}(t)$ changes, with the sole exception of the first torsional mode $b_1(t)$, with the operative conditions reflecting a varying response of the structure. Moreover, the reduced-order torsional dynamics appear to be influenced by the full set of displacement modes, $\bm{\Phi}(z)$. 

\begin{figure}
    \centering
    \includegraphics[width=0.9\linewidth]{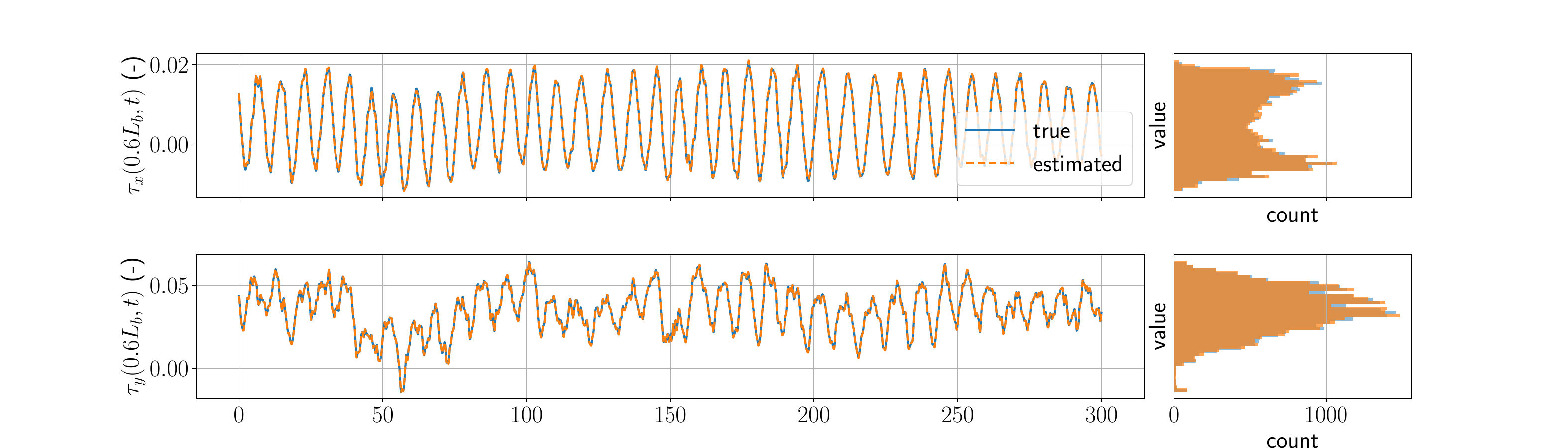}
    \caption{Inferred bending-induced sectional rotations, $\tau_x(z)$ and $\tau_y(z)$, at $z/L_b=0.6$, for $\bar{U}_\infty=10.6$ m/s and TI=$10\%$.}
    \label{fig:tau_xy_rec_10_6}
\end{figure}

\begin{figure}[h!]
\centering
     \begin{subfigure}[b]{0.85\linewidth}
         % \centering
         \includegraphics[width=\linewidth]{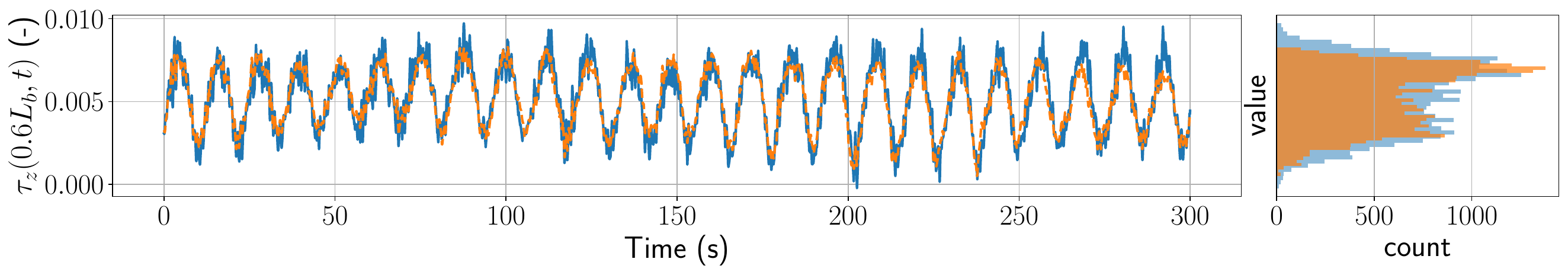}
         \caption{$\bar{U}_\infty =6.2$ m/s.}
         \label{fig:tauzz_6.2}
     \end{subfigure}
     % \hfill
     \begin{subfigure}[b]{0.85\linewidth}
         % \centering
         \includegraphics[width=\linewidth]{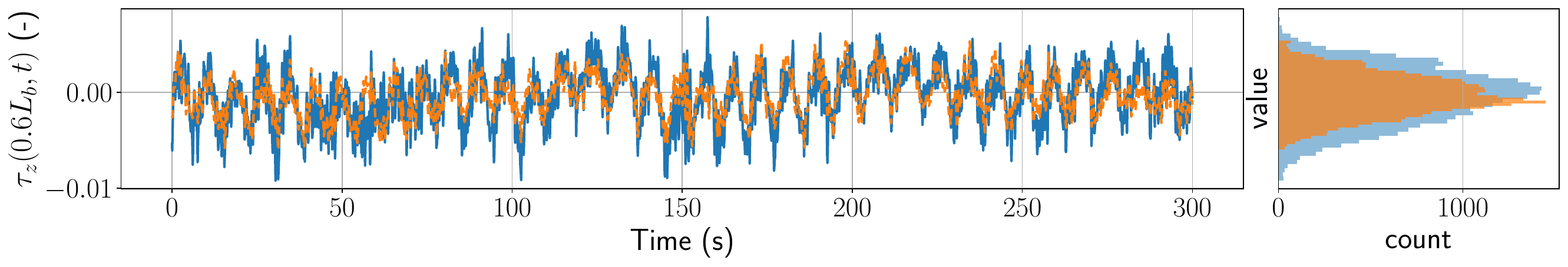}
                  \caption{$\bar{U}_\infty =10.6$ m/s.}
         \label{fig:tauzz_10.6}
     \end{subfigure}
    \begin{subfigure}[b]{0.85\linewidth}
         % \centering
         \includegraphics[width=\linewidth]{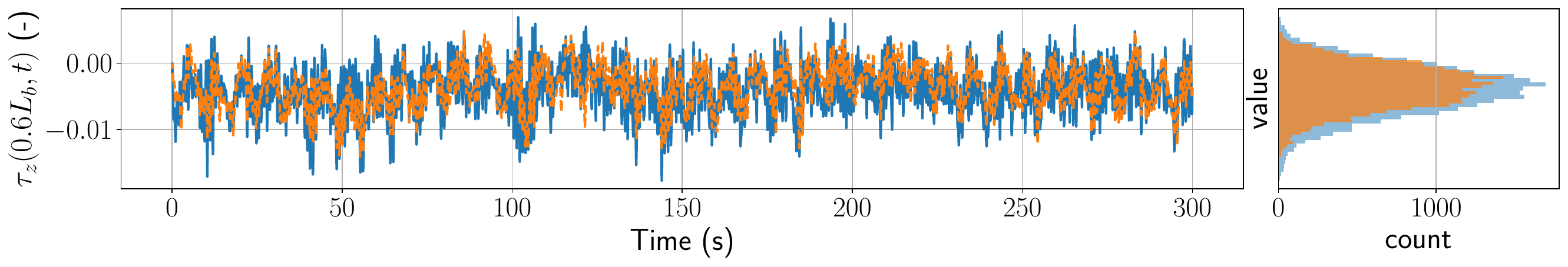}
         \caption{$\bar{U}_\infty =15.0$ m/s.}
         \label{fig:tauzz_15.0}
     \end{subfigure}
    \caption{Inferred pure torsion $\tau_z(z)$ at $z/L_b=0.6$, for $\bar{U}_\infty=10.6$ m/s and TI=$10\%$.}
    \label{fig:tauzz_wspd}
\end{figure}

The torsional POD modes are shown in Figure \ref{fig:torsional_modes_POD}, and the linear map ($\bm{M}$) between this subspace and the one spanned by the POD deflection modes is presented, for different conditions, in Figure \ref{fig:M_wspd}. We reconstruct the full-field with an analogous procedure of what shown in Section §\ref{res_sec_2}, from which we sample an illustrative time series at $z/L_b=0.6$. We note that this evaluation step is carried out on a different time series than the one used to compute $\bm{M}$ (corresponding to a different random seeds for the same condition), to assess the robustness of the method. The bending-induced inferred rotations are shown in Figure \ref{fig:tau_xy_rec_10_6} for $\bar{U}_\infty=10.6$ m/s. $\tau_x(z)$ and $\tau_y(z)$ are directly tied to the beam curvature and, thus, linked with the derivative of the displacement field, explaining the good performances of the linear estimator.
 The reconstruction of the pure torsion $\tau_z(z)$ is shown in Figure \ref{fig:tauzz_wspd} for three different conditions to investigate the capability of the estimator of describing its nonlinear evolution. While exhibiting some evident amplitude mismatches, the linear estimation is able to track the twisting motion of the blade about its longitudinal axis. 

\begin{figure}
    \centering
    \includegraphics[width=0.5\linewidth]{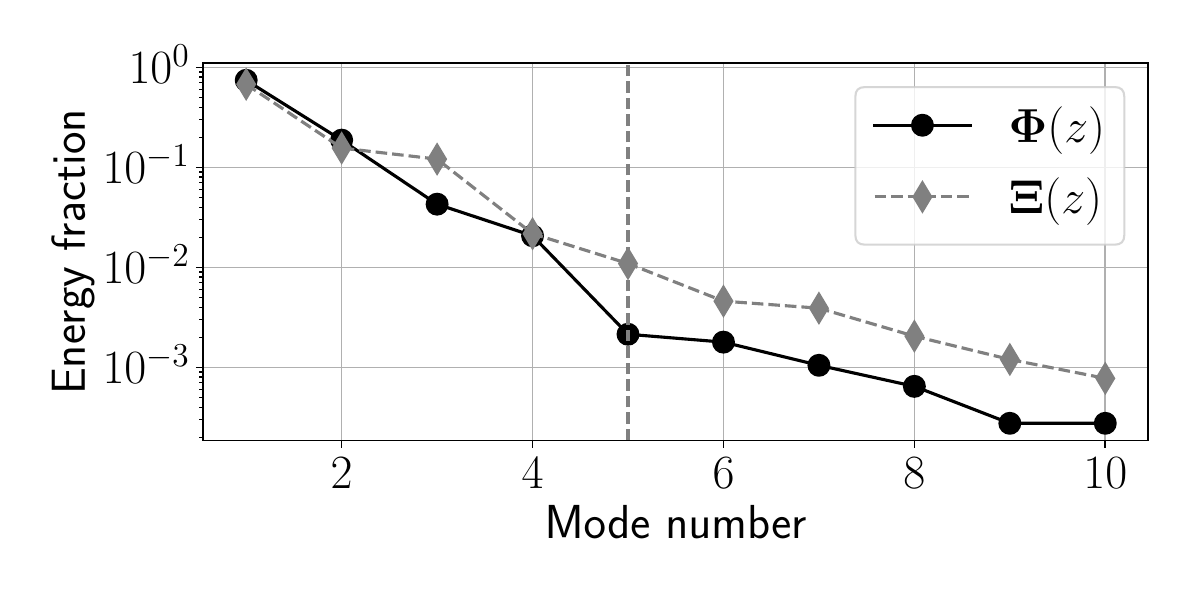}
    \caption{Modal energies of the POD torsional modes $\bm{\Xi}(z)$, compared against the respective deflection ones $\bm{\Phi}(z)$.}
    \label{fig:torsional_eigs}
\end{figure} 
Thus, we first assess the effective dimension of the reduced-order representation of the torsional dynamics. Figure \ref{fig:torsional_eigs} illustrates the modal energies of the torsional modes $\bm{\Xi}(z)= [\bm{\xi}_1(z), \bm{\xi}_2(z), \dots , \bm{\xi}_J(z)]$ and compares it with the deflection modes $\bm{\Phi}(z)$. 
In this case, five modes are found to descriptive of most of the sectional rotations, yielding a reduced-order representation for the torsional motions of an increased rank with respect to the one previously used for the deflections, e.g. $J=N +1$. The variation of the linear map with the operative condition is shown in Figure \ref{fig:M_wspd}.

\nomenclature{LOM}{low order models}
\nomenclature{ROM}{reduced order models}
\nomenclature{LNM}{linear normal modes (eigenmodes)}
\nomenclature{NNM}{nonlinear normal modes}
\nomenclature{POD}{Proper Orthogonal Decomposition}
\nomenclature{POM}{proper orthogonal modes}

\bibliography{Schena_et_al_2024}

\end{document}